\documentclass[12pt,a4paper]{article}
\usepackage{setspace}
\usepackage{mathptmx}

\usepackage[T1]{fontenc}
\usepackage{amsmath}
\usepackage{amssymb}
\usepackage{bm,lscape}
\usepackage{longtable}
\usepackage{mathrsfs,dsfont}
\usepackage{graphicx}
\usepackage{eurosym}
\usepackage{tablefootnote}
\usepackage{footmisc}
\usepackage{subfig}
\usepackage{listings}
\usepackage{adjustbox}
\usepackage{booktabs}
\usepackage{lscape}
 \usepackage[table,xcdraw]{xcolor}
\RequirePackage[colorlinks,citecolor=blue,urlcolor=blue]{hyperref}
\RequirePackage{natbib}
\usepackage[english]{babel}
\usepackage{amsmath,amsthm}
\usepackage{amssymb}
\usepackage{multirow}
\usepackage{bbm}
\usepackage{bm}
\usepackage{hyperref}
\usepackage{mathrsfs}
\usepackage{algorithm}
\usepackage{dsfont}
\usepackage{color}

\usepackage[noend]{algpseudocode}

\makeatletter
\def\BState{\State\hskip-\ALG@thistlm}
\makeatother

\usepackage[margin=1in]{geometry}
\usepackage{lipsum}
\usepackage{ragged2e}

\RaggedRight

\usepackage[markers,tablesfirst,nolists, fighead, tabhead]{endfloat}

\usepackage{endnotes}
\let\footnote=\endnote

\theoremstyle{definition}
\def\bSig\mathbf{\Sigma}

\begin{document}
\doublespacing

\begin{center}
\textmd{\LARGE{\bfseries{{Cointegrated Solutions of Unit-Root VARs: An Extended Representation Theorem}}}}\\

\end{center}
\medskip
\begin{center}
\large{Mario Faliva$^1$ and Maria Grazia Zoia$^2*$}
\end{center}

\bigskip\bigskip
\begin{center}
\small{\noindent $^*$ \textit{Corresponding author}

\noindent $^1$  \textit{Department of Economic Policy, Universit\`a Cattolica del Sacro Cuore, Largo Gemelli 1, 20123, Milano, Italy.  Tel:+390272342948, Fax: +390272342324
\href{mailto:maria.zoia@unicatt.it}{maria.zoia@unicatt.it}}

\noindent $^2$ \textit{Department of Economic Policy, Universit\`a Cattolica del Sacro Cuore, Largo Gemelli 1, 20123, Milano, Italy.  
\href{mailto:mario.faliva@unicatt.it}{mario.faliva@unicatt.it}}}
\end{center}

\begin{center}\textbf{Abstract}\end{center}
This paper establishes an extended representation theorem for unit-root VARs. A specific algebraic technique is devised to recover stationarity from the solution of the model in the form of a cointegrating transformation. 
Closed forms of the results of interest are derived for integrated processes up to the 4-th order. An extension to higher-order processes turns out to be within the reach on an induction argument.

\textit{keywords}:
Unit roots, VAR models, Cointegrated solutions,Stationarity recovering, Parallel sum.


\section*{Introduction}
As is well known, the solution of a unit-root VAR, $\bm{A}(L)\bm{y}_{t}=\bm{\epsilon}_{t}$, crucially rests on the inversion of the matrix polynomial $\bm{A}(L)=\sum_{k=0}^{K}\bm{A}_{k}L^{k}$ and eventually of the isomorphic  matrix $\bm{A}(z)$ in the complex variable $z$ about the unit root $z$=1. The solution takes the form of an integrated stochastic process, where stationarity can be recovered by a cointegrating transformation.\\
The inversion of a matrix polynomial plays a crucial role in the representation theory of linear processes. 
The seminal and best known contribution to this topic is the so-called Granger representation theorem by ~\cite{granger1981some}, ~\cite{granger1983co}  and ~\cite{engle1987cointegration} , that addressed the issue of the inversion of a matrix polynomial inherent in the MA representation of an integrated process and derived the so-called error-correction model. Since then, a stream of research and contributions have been registered on this issue, leading to a specialized literature. Among them, ~\cite{phillips1991optimal}, ~\cite{phillips1990statistical}, ~\cite{sims1990inference}, ~\cite{stock1993simple} who worked out the triangular representation, and ~\cite{engle1991ncointegrated},  ~\cite{haldrup1998representations} who introduced the use of the Smith-MacMillan form of $A(L)$. Johansen, ~\cite{johansen1985mathematical}, ~\cite{johansen1991estimation}, ~\cite{johansen1992representation} and  ~\cite{johansen1996likelihood}, developed the Granger's representation theorem in the context of integrated VAR models and established the necessary and sufficient conditions for the occurrence of first and second-order integrated processes (see ~\cite{franchi2019general} for a detailed discussion of the Granger representation theorem history).\\
First ~\cite{schumacher1991system} pointed out that Johansen's $I(1)$ conditions could be restated in terms of existence of a simple pole in the inverse of the VAR autoregressive polynomial. Several authors have investigated the problem of the inversion of a matrix polynomial about a pole (see e.g.,~\cite{avrachenkov2001inversion},~\cite{faliva2002partitioned},~\cite{faliva2003new},~\cite{faliva2008dynamic},~\cite{faliva2011inversion},~\cite{langenhop},~\cite{franchi2019general}) ever since. This topic has been also recently revisited from the Hilbert-space standpoint via the theory of functional time series by ~\cite{beare2017cointegrated},~\cite{beare2019representation}   and  ~\cite{franchi2019cointegration}. \\
This paper develops an approach to unit-root VAR models which crucially hinges on the Laurent expansion of the inverse of the isomorphic matrix polynomial $\bm{A(z)}$. This eventually leads to determine the VAR solution, corresponding to the order of the pole of $\bm{A(z)}^{-1}$, along with its integration and cointegration properties. Closed-form expressions of the results of interest are provided for VAR models with (co)integrated solutions up to the 4-th order. By an induction argument, the analysis can be further advanced to cover processes of higher orders.

The paper develops in a twofold way. Once an extended unit-root VAR representation theorem is stated in Section 1, the paper switches to set up the required analytical apparatus, which is provided in Sections 2 and 3. Here, closed-form expressions of the principal-part matrices in the Laurent expansion of $\bm{A}(z)^{-1}$ about a unit root are worked out. The key issue of recovering stationarity, via a linear transformation of $\bm{A}(z)^{-1}$ which annihilates the principal part, is successfully faced. 

The algebraic set-up of the paper pivots around the twin equalities $\bm{A}^{-1}(z)\bm{A}(z)=\textbf{I}$ and $\bm{A}(z)\bm{A}^{-1}(z)=\textbf{I}$ which hold true in a deleted neighbourhood of $z=1$. The twin equation systems 
that arise from the said equalities, allow to obtain informative closed-form expressions of the principal-part matrices. It can be shown that, besides the leading matrix, all the other principal-part matrices obey a regular scheme, as they can be expressed as a sum of two components: a term whose representation does not vary with the order of the pole and another which changes according to the latter. For a  principal part matrix, $\bm{N}_{j}$, which weights the power $(z-1)^{-j}$, the former term turns out to be a linear combination of all the principal-part matrices, $\bm{N}_{i}$, weighting (negative) higher order powers of $(z-1)$, namely the matrices $\bm{N}_{i}$ weighting $(z-1)^{-i}$, $i=j+1,..,m$, where $m$ is the order of the pole. 
The coefficient matrices of the said linear combination play a crucial role in the cointegration analysis.\\ 
The multiplicity of the pole is determined and  analyticity at $z=1$ is recovered via a transformation  $\bm{P}\bm{A}^{-1}(z)$, where $\bm{P}$ is a projection operator of the principal-part on the null space. Use  of the notion of parallel sum of matrices is made to find out the cointegrating relationships and their rank. When the order of the pole is multiple, the problem of annihilating the principal-part is solved by means of more operators that jointly meet the target. To this end, the role of parallel sum of matrices proves effective to combine all the projectors needed to annihilate the principal-part. 
Having established the results we needed, the paper turns back to the econometric side of the problem and provides in Section 4 the proof of the unit-root VAR theorem of Section 1. 
From an econometric standpoint, the value added by the paper is attributable to the approach to determine the unit-root VAR solutions and the innovative stationarity recovery technique. In the paper such topics are fully developed for VAR models with solutions integrated up to the $4^{th}$ order. Thus the paper crosses the virtual threshold of I(2) processes and clears the way to tackle cointegrated processes of higher order by induction. 
The paper is organized as follows. Section~\ref{sec:1sec} formulates the extended theorem which provides the solution of unit-root VARs together with its integration and co-integration properties. Sections~\ref{sec:2sec} and ~\ref{sec:3sec} work out the analytical premise and the algebraic results, demanded to establish the main theorem of Section 1. Section 4 gives the proof of the said theorem.  Section 5 provides some concluding remarks. Two appendices complete the paper, the former is devoted to parallel sums of matrices and the latter derives the results and formulas demanded by Theorem 3.1. \\

\section{A basic theorem for VAR models with unit roots}{\label{sec:1sec}}
In this section an extended representation theorem is established for unit-root VAR models whose solutions are integrated processes up to the $4^{th}$ order and stationarity recovering via cointegration is thoroughly investigated. The latter  
is not only interesting in itself but plays a crucial role in economic analysis insofar as it offers a key to the interpretation of the long-run dynamics inherent in economic phenomena under investigation (see e.g., ~\cite{banerjee1993co} ). As the theorem demands an $ad \enspace hoc$ analytic apparatus, 
its proof is postponed until the intended algebraic toolkit is made available in the newt two sections. \\
Let us now state the following

\noindent \textbf{Theorem 1.1}\\
\noindent \textit{Consider the VAR model:}
\begin{equation}\label{eq:1eq}
\bf{A}(L)\bm{y}_{t}=\bm{\varepsilon}_{t}, \,\,\,\,\,\, \underset{(n,1)}{\bm{\varepsilon}_{t}}\sim WN_{n}(\bm{0},\bm{\Sigma})
\end{equation}
\textit{where}
\begin{equation}\label{eq:2eq}
\bm{A}(L)=\bm{I}_{n}-\sum_{k=1}^{K}\bm{A}_{k}L^{k}.
\end{equation}
\textit{Let $z$=1 be a root of multiplicity $\mu$ of the characteristic polynomial $\bm{A}(z)$, with the other roots lying outside the unit circle.
Then, the solution $\bm{y}_{t}$ of \eqref{eq:1eq} is an integrated $(\bm{I})$ and co-integrated process, that is}
\begin{equation}\label{eq:2seq}
\bm{y}_{t}\sim \bm{I}(m)
\end{equation}
\begin{equation}\label{eq:3eq}
\bm{P}_{m}\bm{y}_{t}\sim \bm{I}(0)
\end{equation}
\textit{where $m$ ($m \le \mu$) is the least positive integer for which the matrix}
\begin{equation}\label{eq:4eq}
\bm{K}_m=(\prod_{i=0}^{m-1}\bm{B}_{i\bot})\bm{A}^{[m]}(\prod_{i=0}^{m-1}\bm{C}_{i\bot})
\end{equation}
\textit{is non-singular. The matrices $\bm{B}_{j}$, $\bm{C}_{j}$ arise from the rank factorizations }
\begin{equation}\label{eq:5eq}
\bm{A}(1)=\bm{B}_{0}\bm{C}_{0}
\end{equation}
\begin{equation}\label{eq:6eq}
\bm{K}_i=\bm{B}_{i}\bm{C}_{i}^{'}, \enspace 1\leq i < m,
\end{equation}
\textit{ 
where the subscript $\bot$ stands for orthogonal complement, and} 
\begin{equation}\label{eq:7eq}
\bm{A}^{[m]}=\bm{A}^{(1)} \enspace \enspace \enspace if\enspace m=1
\end{equation}
\begin{equation}\label{eq:8eq}
\bm{A}^{[m]}=\frac{1}{2}\bm{A}^{(2)}-\widetilde{\bm{A}}^{(2)}, \enspace  \enspace \enspace \enspace \enspace \widetilde{\bm{A}}^{(2)}=\bm{A}^{(1)}\bm{A}^{+}\bm{A}^{(1)}\enspace \enspace  \enspace \enspace \enspace if \enspace m=2
\end{equation}
\begin{equation}\label{eq:9eq}
\bm{A}^{[m]}=\frac{1}{6}\bm{A}^{(3)}-\widetilde{\bm{A}}^{(3)}, \enspace  \enspace \enspace  \widetilde{\bm{A}}^{(3)}=\begin{bmatrix}\bm{A}^{(1)}\bm{A}^{+},& \bm{A}^{[2]}\end{bmatrix}\begin{bmatrix}
\bm{A}^{(1)} & \bm{I}\\
\bm{I} & \bm{\Theta}_1
\end{bmatrix}\begin{bmatrix}
\bm{A}^{+}\bm{A}^{(1)} \\
\bm{A}^{[2]} 
\end{bmatrix}
\enspace  \enspace if\enspace m=3
\end{equation}
\begin{center}
$\bm{A}^{[m]}=\frac{1}{24}\bm{A}^{(4)}-\widetilde{\bm{A}}^{(4)},$    
\end{center}
\begin{equation}\label{eq:10eq}
\widetilde{\bm{A}}^{(4)}
=\begin{bmatrix}\bm{A}^{(1)}\bm{A}^{+},&\bm{A}^{[2]},\bm{A}^{[3]}\end{bmatrix}
\begin{bmatrix}
\frac{1}{2}\bm{A}^{(2)} &\bm{A}^{[2]}\bm{\Theta}_{1}+ \bm{A}^{(1)}\bm{A}^{+}&\bm{I}\\
\bm{A}^{+}\bm{A}^{(1)}+\bm{\Theta}_{1}\bm{A}^{[2]}&
\bm{A}^{+}\bm{A}^{(1)}+\bm{\Theta}_{1}\bm{A}^{[2]}& \bm{A}^{+}+\bm{\Theta}_{1}\bm{A}^{[2]}\bm{\Theta}_{1}&\bm{\Theta}_1\\
\bm{I} & \bm{\Theta}_1 & \bm{\Theta}_2
\end{bmatrix}
\begin{bmatrix}
\bm{A}^{+}\bm{A}^{(1)} \\
\bm{A}^{[2]} \\
\bm{A}^{[3]}
\end{bmatrix} \enspace \enspace \enspace if\enspace m=4  
\end{equation}
\textit{Here $\bm{A}^{+}$ denotes the Moore-Penrose generalized inverse of $\bm{A}$, }
\begin{equation}\label{eq:11eq}
\bm{A}^{(k)}=\frac{	\partial^{k}\bm{A}(z)}{	\partial z^k}\bigm|_{z=1}, \enspace \enspace\enspace \enspace \bm{A}^{(0)}=\bm{A}(1)=\bm{A}
\end{equation}
\begin{equation}\label{eq:12eq}
\bm{\Theta}_{1}=\bm{C}_{0\bot}\bm{K}^{+}_{1}\bm{B}^{'}_{0\bot}
\end{equation}
\begin{equation}\label{eq:13eq}
\bm{\Theta}_{2}=\bm{C}_{0\bot}\bm{C}_{1\bot}\bm{K}^{+}_{2}\bm{B}^{'}_{1\bot}\bm{B}^{'}_{0\bot}
\end{equation}
\textit{The cointegration matrices in \eqref{eq:3eq} are} 
\begin{equation}\label{eq:14eq}
\bm{P}_m=(\bm{C}^{'}_{0})^{+}\bm{C}^{'}_{0} \enspace\enspace\enspace \enspace\enspace\enspace if \enspace m=1
\end{equation}
\begin{equation}\label{eq:15eq}
\bm{P}_m= -\begin{bmatrix}\bm{0}, & \bm{0}, &\bm{I}\end{bmatrix}
\begin{bmatrix} 2\bm{P}_1 & \bm{0} &\bm{I}\\
\bm{0} & 2\bm{\Pi}_2 &\bm{I} \\
\bm{I} & \bm{I} &\bm{0}\end{bmatrix}\begin{bmatrix}
\bm{0}\\ \bm{0}\\\bm{I}
\end{bmatrix}
\enspace\enspace\enspace \enspace\enspace\enspace if \enspace m=2
\end{equation}
\begin{equation}\label{eq:16eq}
\bm{P}_m=-\begin{bmatrix}\bm{0}, & \bm{0}, &\bm{I}\end{bmatrix}
\begin{bmatrix} 2\bm{P}_1 & \bm{0} &\bm{I}\\
\bm{0} & 2\bm{\Pi}_3 &\bm{I} \\
\bm{I} & \bm{I} &\bm{0}\end{bmatrix}\begin{bmatrix}
\bm{0}\\ \bm{0}\\\bm{I}
\end{bmatrix}
\enspace\enspace\enspace \enspace\enspace\enspace if \enspace m=3
\end{equation}
\begin{equation}\label{eq:17eq}
\bm{P}_m=-\begin{bmatrix}\bm{0}, & \bm{0}, &\bm{I}\end{bmatrix}
\begin{bmatrix} 2\bm{P}_1 & \bm{0} &\bm{I}\\
\bm{0} & 2\bm{\Pi}_4 &\bm{I} \\
\bm{I} & \bm{I} &\bm{0}\end{bmatrix}\begin{bmatrix}
\bm{0}\\ \bm{0}\\\bm{I}
\end{bmatrix}
\enspace\enspace\enspace \enspace\enspace\enspace if \enspace m=4
\end{equation}
\textit{where}
\begin{equation}\label{eq:18eq}
\bm{\Pi}_2=(\bm{A}^{+}\bm{A}^{(1)}\bm{C}_{0\bot}\bm{C}_{1\bot})^{\top}
\end{equation}
\begin{equation}\label{eq:19eq}
\bm{\Pi}_3=2((\bm{A}^{+}\bm{A}^{(1)}\bm{C}_{0\bot})^{\top}:(\bm{A}^{+}\bm{A}^{[2]}\bm{C}_{0\bot}\bm{C}_{1\bot}\bm{C}_{2\bot})^{\top})
\end{equation}
\begin{equation}\label{eq:20eq}
\bm{\Pi}_4=2(\bm{\Pi}_{3,4}:(\bm{A}^{+}\bm{A}^{[2]}\bm{C}_{0\bot}\bm{C}_{1\bot}\bm{C}_{2\bot}\bm{C}_{3\bot})^{\top})
\end{equation}
\textit{with}
\begin{equation}\label{eq:21eq}
\bm{\Pi}_{3,4}=2((\bm{A}^{+}\bm{A}^{(1)}\bm{C}_{0\bot})^{\top}:(\bm{A}^{+}\bm{A}^{[2]}\bm{C}_{0\bot})^{\top})
\end{equation}\\
\textit{Here $\bm{X: Z=X(X+Z)^{+}Z}$ denotes the parallel sum of $\bm{X}$ and $\bm{Z}$, and $\bm{G^{\top}=I-GG^{+}}$. The cointegration ranks are}
\begin{equation}\label{eq:22eq}
r(\bm{P}_m)=r(\bm{C}_0) \enspace\enspace\enspace\enspace\enspace\enspace
if \enspace m=1
\end{equation}
\begin{equation}\label{eq:23eq}
r(\bm{P}_m)=r(\bm{C}_0)-r(\bm{A}^{+}\bm{A}^{(1)}\bm{C}_{0\bot}\bm{C}_{1\bot})
\enspace\enspace\enspace\enspace\enspace\enspace
if \enspace m=2
\end{equation}
\begin{equation}\label{eq:24eq}
r(\bm{P}_m)=r(\bm{C}_0)-r(\begin{bmatrix}
\bm{A}^{+}\bm{A}^{(1)}\bm{C}_{0\bot}, &
\bm{A}^{+}\bm{A}^{[2]}\bm{C}_{0\bot}\bm{C}_{1\bot}\bm{C}_{2\bot}
\end{bmatrix})
\enspace\enspace\enspace\enspace\enspace\enspace if \enspace m=3
\end{equation}
\begin{equation}\label{eq:424eq}
 r(\bm{P}_m)=r(\bm{C}_0)-r(\begin{bmatrix}
\bm{A}^{+}\bm{A}^{(1)}\bm{C}_{0\bot}, &\bm{A}^{+}\bm{A}^{[2]}\bm{C}_{0\bot}
\end{bmatrix}) 
+r(\bm{\Xi})-r([\bm{\Xi},\bm{\Gamma}])
\enspace\enspace\enspace\enspace\enspace if \enspace m=4
\end{equation}   
\textit{with $\bm{\Gamma}\bm{\Gamma}^{+}=\bm{\Pi}_{3,4}$ and $\bm{\Xi}\bm{\Xi}^{+}=(\bm{A}^{+}\bm{A}^{[3]}\bm{C}_{0\bot}\bm{C}_{1\bot}\bm{C}_{2\bot}\bm{C}_{3\bot})^{\top})$} \\
\textbf{Proof}\\
Go to Section~\ref{sec:4sec}
\begin{flushright}
$\square$
\end{flushright} 
Several facts on the inversion of a matrix polynomial about a pole must be established before proving the theorem. This is done in the following two Sections.

\section{Matrix Polynomials and their inversion about a pole}{\label{sec:2sec}}
 As the solution of a unit-root VAR model, $\bm{A}(z)\textbf{y}_{t}=\bm{\epsilon}_{t}$, crucially rests on the operator $\bm{A}^{-1}(L)$ (see e.g., ~\cite{faliva2008dynamic}, Sections 2.3 and 2.9 ) and the algebra of matrix polynomials in the lag operator $L$ and in a complex variable $z$ are isomorphic (~\cite{dhrymes1971distributed}), let us address the issue of the inversion of $\bm{A}(z)$ about a pole, $z=z_{o}$, with $z_{o}=1$ for our purposes. \\
 Starting from the two basic equalities $\bm{A}^{-1}(z)\bm{A}(z)=\textbf{I}$ and $\bm{A}(z)\bm{A}^{-1}(z)=\textbf{I}$, which hold true in a deleted neighbourhood of a pole, we derive two equation systems which allow eventually to work out closed-form expressions for the coefficient matrices of the principal part  of $\bm{A}^{-1}(z)$. \\
 Let
 \begin{equation}\label{eq:26eq}
 \bm{A}(z)=\sum_{k=0}^{K}\bm{A}_k z^k, \enspace\enspace\enspace\enspace\underset{(n, n)}{\bm{A}_k} \ne \bm{0}
\end{equation}
 be a matrix polynomial of order $n$ and degree $K$ and $z_0$ denotes a root of
  \begin{equation}\label{eq:27eq}
  det(\bm{A}(z))=0
\end{equation}
Expanding $\bm{A}(z)$ about $z=z_0$ yields
 \begin{equation}\label{eq:28eq}
 \bm{A}(z)=\sum_{k=0}^{K}\frac{1}{k!}\bm{A}^{(k)}(z-z_0)^k
\end{equation}
As the matrix function $\bm{A}^{-1}(z)$ is analytic through the $z$-plane except for the zeros, $z_0$, of $det\bm{A}(z)$, $\bm{A}^{-1}(z)$, the following  Laurent expansion  
\begin{equation}\label{eq:29eq}
 \bm{A}^{-1}(z)=\sum_{j=-m}^{\infty}\bm{N}_j(z-z_0)^j=\sum_{j=-m}^{-1}\bm{N}_j(z-z_0)^j + \bm{M}(z)
\end{equation}
holds in a deleted neighborhood of the pole located at $z=z_0$ (see, e.g.,~\cite{faliva2008dynamic}). Here $m$ is the order of the pole. The first term on the right-hand side of \eqref{eq:29eq} is the principal part, while $\bm{M}(z)=\sum_{j=0}^{\infty}\bm{N}_{j}(z-z_0)^{j}$ is the regular part. 

In a deleted neighbourhood of $z=z_0$, the product of the right hand sides of \eqref{eq:29eq} and \eqref{eq:28eq} yields the equalities
\[
\bm{I}_n=\sum_{j=-m}^{\infty}\bm{N}_j(z-z_0)^j  \left(\sum_{k=0}^{K}\frac{1}{k!}\bm{A}^{(k)}(z-z_0)^k \right)
=\sum_{k=-m}^{\infty}\left(\sum_{j=0}^{m+k}\frac{1}{j!}\bm{N}_{k-j}\bm{A}^{(j)}\right)(z-z_0)^k=
\]
\begin{equation}\label{eq:30eq}
=\sum_{h=0}^{\infty}\left(\sum_{j=0}^{h}\frac{1}{j!}\bm{N}_{h-m-j}\bm{A}^{(j)}\right)(z-z_0)^{h-m}
\end{equation}
In turn, reversing the order of multiplication yields 
\begin{center}
$\bm{I}_n=\left(\sum_{k=0}^{K}\frac{1}{k!}\bm{A}^{(k)}(z-z_0)^k \right)\sum_{j=-m}^{\infty}\bm{N}_j(z-z_0)^j=\sum_{k=-m}^{\infty}\left(\sum_{j=-m}^{m+k}\frac{1}{j!}\bm{A}^{(j)}\bm{N}_{k-j}\right)(z-z_0)^k=$
\end{center} 
\begin{equation}\label{eq:31eq}
=\sum_{h=0}^{\infty}\left(\sum_{j=0}^{h}\frac{1}{j!}\bm{A}^{(j)}\bm{N}_{h-m-j}\right)(z-z_0)^{h-m} 
\end{equation}
\vspace{5mm}
The coefficient matrices in the right-hand side of \eqref{eq:30eq} associated with negative powers of $(z-z_0)$ are null matrices, whereas  the matrix associated with  $(z-z_0)^0$ is the identity matrix, that is    
\begin{equation}\label{eq:32eq}
\sum_{j=0}^{h}\frac{1}{j!}\bm{N}_{-m+k-j}\bm{A}^{(j)}=\left\{\begin{array}{ll}
\bm{0}\enspace\enspace\enspace\enspace\enspace if \enspace\enspace\enspace 0\leq h\leq m-1 \\
\bm{I}_n\enspace\enspace\enspace\enspace if
\enspace\enspace\enspace h=m\\
\end{array}\right.
\end{equation}
The same argument applies to equation \eqref{eq:31eq} and
\begin{equation} \label{eq:33eq}
 \sum_{j=0}^{h}\frac{1}{j!}\bm{A}^{(j)}\bm{N}_{-m+k-j}=\left\{\begin{array}{ll}
\bm{0}\enspace\enspace\enspace\enspace\enspace if \enspace\enspace\enspace 0\leq h\leq m-1 \\
\bm{I}_n\enspace\enspace\enspace\enspace if
\enspace\enspace\enspace h=m\\
\end{array}\right.
\end{equation}
follows accordingly.                  
\\Hereafter, the analysis is concerned with unit roots, $z_0$=1 , which entails that $\bm{A}(1)=\bm{A}$ is a  singular matrix. \\
\section{Pole order, Laurent expansion and analyticity recovering}{\label{sec:3sec}}
In this section we establish under which conditions  $\bm{A}^{-1}(z)$ has a pole of order $m$ at $z=z_0=1$ , derive closed form expressions of the principal-part matrices and determine linear functions of $\bm{A}^{-1}(z)$ which are analytic at $z=1$. \\
First of all, we address the issue of finding informative expressions of the principal-part coefficient matrices in $\bm{A}^{-1}(z)$. This is done in Theorem 3.1. Here, both a basic result on the leading principal-part matrix and useful representations of the non-leading ones are provided for multiple poles. These representations turn out to be the resultant of two terms: a term whose structure is maintained as the pole order changes and a term which is peculiar to the multiplicity of the  pole and vanishes if the pole is simple. In order to unburden the exposition,  the working out of formulas is left to an Appendix (B). \\ 
Afterwords, Theorem 3.2 determines the order of the pole of $\bm{A}^{-1}(z)$, gives closed-form representations of the principal-part leading matrix of the Laurent expansion of $\bm{A}^{-1}(z)$ about $z=1$ and ascertains which linear transformations recover analycity at $z=1$.
\\
\textbf{Theorem 3.1}
\\
\textit{Let $z=z_0$=1 be a pole of $\bm{A}^{-1}(z)$. Then, the following holds }\\
\begin{enumerate}
\item \textit{The leading principal-part matrix $\bm{N}_{-m}$  of the Laurent expansion of $\bm{A}^{-1}(z)$ about $z=1$ has the representation} 
\begin{equation}\label{eq:34eq}
\bm{N}_{-m}=\bm{C}_{0\bot}\bm{Z}_{m}\bm{B}_{0\bot}^{'}
\end{equation}
\textit{for some $\bm{Z}_{m}$.}
\item \textit{The non-leading principal-part matrices $\bm{N}_{-m+\theta}$, $ 1\leq \theta < m \leq 4$, is composed of two terms,}
\begin{equation}\label{eq:R35eq}
\bm{N}_{-m+\theta}=\bm{\Lambda}_{\theta}+ \bm{\Lambda}_{m,\theta}
\end{equation}
\end{enumerate}
\noindent
\textit{The first term, $\bm{\Lambda}_{\theta}$, has a structure which depends only on ${\theta}$ and plays a crucial role in determining the linear transformations which recovers stationary from the VAR solution. The matrix $\bm{\Lambda}_{\theta}$ is the sum of the discrete convolutions of $\frac{1}{j!}\bm{A}^{+}\bm{A}^{j}$ and $-\bm{N}_{-m+\theta-j}$, of -$\bm{N}_{-m+\theta-j}$ and $\frac{1}{j!}\bm{A}^{+}\bm{A}^{j}$,and of  $\bm{A}^{+}\bm{A}\bm{N}_{-m+\theta-j}$, and $\frac{1}{j!}\bm{A}^{j}\bm{A}^{+}$, respectively, that is } \begin{equation*}
\bm{\Lambda}_{\theta}=-\bm{A}^{+}(\sum_{j=1}^{\theta}\frac{1}{j!}\bm{A}^{(j)}\bm{N}_{-m+\theta-j})-(\sum_{j=1}^{\theta}\bm{N}_{-m+\theta-j}\frac{1}{j!}\bm{A}^{(j)})\bm{A}^{+}+
\end{equation*}
\begin{equation}\label{eq:RR35eq}
+\bm{A}^{+}\bm{A}\sum_{j=1}^{\theta-1}\bm{N}_{-m+\theta-j}\frac{1}{j!}\bm{A}^{(j)}\bm{A}^{+}
\end{equation}
\textit{The second term, $\bm{\Lambda}_{m,\theta}$ has a structure which depends on both $\theta$ and the order, $m$, of the pole order. In particular, the following holds}
\begin{equation}\label{eq:36eq}
\bm{\Lambda}_{m,1}=
\begin{cases}
\bm{0} \enspace \enspace \enspace \enspace  \enspace\enspace \enspace\enspace  \enspace \enspace \enspace \enspace  \enspace\enspace \enspace \enspace \enspace  \enspace\enspace \enspace\enspace  \enspace \enspace \enspace \enspace  \enspace \enspace \enspace \enspace \enspace  \enspace\enspace \enspace\enspace  \enspace \enspace \enspace \enspace  \enspace  \enspace\text{if} \enspace \enspace    m<2, \\
\bm{C}_{0, \bot}\bm{S}_{1,m-1}\bm{B}_{0, \bot}'
\enspace \enspace \enspace \enspace  \enspace\enspace \enspace \enspace \enspace \enspace  \enspace\enspace \enspace \enspace \enspace  \enspace\enspace \enspace\enspace  \enspace \enspace \enspace \enspace  \enspace \enspace \enspace \enspace\text{if} \enspace \enspace    m=2, \\
-\sum_{j=1}^{m-2}(\bm{\Theta}_{j}\bm{A}^{[j+1]}\bm{N}_{-m}+\bm{N}_{-m}\bm{A}^{[j+1]}\bm{\Theta}_{j})+\\
+(\prod_{j=0}^{m-2}{\bm{C}_{j\bot}})\bm{S}_{1,m-1}(\prod_{j=0}^{m-2}{\bm{B}_{m-2-j\bot}^{'}})\enspace \enspace  \enspace\enspace \enspace \enspace\enspace  \enspace  \text{if}\enspace\enspace m>2,
\end{cases}
\end{equation}
\textit{for $\theta=1$ and for some $\bm{S}_{1,m-1}$}.
\textit{Here $\bm{\Theta}_{1}$ and $\bm{\Theta}_{2}$ are the matrices specified in \eqref{eq:12eq} and \eqref{eq:13eq}, respectively, and}
\begin{equation}\label{eq:38eq}
\bm{K}_{1}=\bm{B}_{0\bot}^{'}\bm{A}^{(1)}\bm{C}_{0\bot}   
\end{equation}
\begin{equation}\label{eq:39eq}
\bm{K}_{2}=\bm{B}_{1\bot}^{'}\bm{B}_{0\bot}^{'}\bm{A}^{[2]}\bm{C}_{0\bot}\bm{C}_{1\bot}  
\end{equation}
\begin{equation}\label{eq:36eqz}
\bm{\Lambda}_{m,2}=
\begin{cases}
\bm{0} \enspace \enspace \enspace \enspace  \enspace\enspace \enspace\enspace  \enspace \enspace \enspace \enspace  \enspace\enspace \enspace \enspace \enspace  \enspace\enspace \enspace\enspace  \enspace \enspace \enspace \enspace  \enspace \enspace \enspace \enspace \enspace  \enspace\enspace \enspace\enspace  \enspace \enspace \enspace \enspace  \enspace  \enspace\text{if} \enspace \enspace    m<3, \\
\bm{C}_{0, \bot}\bm{S}_{2,m-1}\bm{B}_{0, \bot}'
\enspace \enspace \enspace \enspace  \enspace\enspace \enspace \enspace \enspace \enspace  \enspace\enspace \enspace \enspace \enspace  \enspace\enspace \enspace\enspace  \enspace \enspace \enspace \enspace  \enspace \enspace \enspace \enspace\text{if} \enspace \enspace    m=3, \\
-\bm{\Theta}_{1}\bm{A}^{[2]}\bm{N}_{-m+1}(\bm{B}^{'}_{0\bot})^{+}\bm{K}^{\top}_{1}\bm{B}^{'}_{0\bot}-\bm{\Theta}_{1}\dot{\bm{A}}^{[3]}\bm{N}_{-m}+\\
-\bm{C}_{0\bot}\bm{C}^{+}_{0\bot}\bm{N}_{-m+1}\bm{A}^{[2]}\bm{\Theta}_{1}-\bm{N}_{-m}\breve{\bm{A}}^{[3]}\bm{\Theta}_{1}+\\
+\bm{C}_{0\bot}\bm{C}_{1\bot}\bm{S}_{2,m-1}\bm{B}^{'}_{1\bot}\bm{B}^{'}_{0\bot}\enspace \enspace \enspace\enspace  \enspace \enspace \enspace \enspace \enspace  \enspace \enspace \enspace \enspace \enspace  \enspace \enspace \enspace \enspace  \enspace  \enspace\enspace \enspace\text{if} \enspace \enspace \enspace   m>3,
\end{cases}
\end{equation}
\textit{for $\theta=2$ and for some $\bm{S}_{2,m-1}$. Here $\dot{\bm{A}}^{[3]}$ and $\breve{\bm{A}}^{[3]}$ denote the matrices \eqref{eq:156eq} and \eqref{eq:157eq} in Appendix B}.
\\\textbf{Proof}\\
The proof follows from the representations of the principal-part matrices $\bm{N}_{m+1}$, $\bm{N}_{m+2}$ and $\bm{N}_{m+3}$ of formulas \eqref{eq:148eq}, \eqref{eq:152eq} and \eqref{eq:159eq} (Appendix B). In particular, formula \eqref{eq:36eq} rests on  \eqref{eq:148eq} together with \eqref{eq:148eqzl} and \eqref{eq:148eqzt} in the said Appendix, while formula \eqref{eq:36eqz} hinges on \eqref{eq:152eq} and  \eqref{eq:158eq}).
\begin{flushright}
$\square$
\end{flushright}
At this point we can derive the results we are mostly interested in. To this end, next theorem establishes the order of the pole of $\bm{A}(z)^{-1}$ by a determinantal criterion, gives the closed-form representations of the leading matrices of the principal-part for poles of order $1\leq m\leq 4$, determines which linear forms $\bm{P}_{m}\bm{A}(z)^{-1}=\bm{P}_{m}\bm{M}(z)$ recover analyticity at $z=1$ and eventually ascertains the ranks of the matrices $\bm{P}_{m}$ which are orthogonal to the principal-part of $\bm{A}(z)^{-1}$. \\
It is worth noting that the propositions of the theorem which follows have an econometric counterpart of prominent interest, insofar as they clear the way, thanks to the isomorphism of algebras in $L$ and $z$, to the cointegration analysis in unit-root VAR models. Indeed, the order of the pole of $\bm{A}(z)^{-1}$ determines the integration order of the solution of $\bm{A}(L)\bm{y}_{t}=\boldsymbol{\epsilon}_{t}$ and the analyticity of $\bm{P}_{m}\bm{A}(z)^{-1}$ pairs off with the stationarity of the linear transformations $\bm{P}_{m}\bm{y}_{t}$, which shed light into the otherwise hidden long-run relationships of an economic system. \\
\textbf{Theorem 3.2}\\
\textit{Let $\bm{A}(z)$ have a possibly repeated unit root, $z$=1, and  $\bm{A}=\bm{B}_0\bm{C}_0^{'}$ be a rank-factorization of the singular matrix $\bm{A}(1)=\bm{A}$. Then, the following statements hold}\vspace{5mm}
\begin{enumerate}
\item \textit{the unit root $z=1$ is a simple pole of $\bm{A}^{-1}(z)$, that is $m=1$, if}  
\begin{equation}\label{eq:44eq}
det\bm{K}_1\neq 0    
\end{equation}
\textit{where $\bm{K}_1$ is defined as in \eqref{eq:38eq}.\\
Under \eqref{eq:44eq}, the matrix $\bm{A}^{-1}(z)$ has the Laurent expansion \eqref{eq:29eq} about $z=1$, with $m=1$ and}
\begin{equation}\label{eq:45eq}
\bm{N}_{-1}=\bm{C}_{0\bot}(\bm{B}_{0\bot}^{'}\bm{A}^{(1)}\bm{C}_{0\bot})^{-1}\bm{B}_{0\bot}^{'}
\end{equation}
\textit{as residue matrix.\\
The matrix function $\bm{P}_{1}\bm{A}^{-1}(z)$ is analytic at $z=1$ for}
\begin{equation}\label{eq:46eq}
\bm{P}_{1}=(\bm{C}_{0}^{'})^{+}\bm{C}_{0}^{'}
\end{equation}
\item \textit{The unit root $z=1$ is a double pole of $\bm{A}^{-1}(z)$, that is $m=2$, if}
\begin{equation}\label{eq:47eq}
det\bm{K}_{1}=0
\end{equation}
\begin{equation}\label{eq:48eq}
det\bm{K}_{2}\neq 0
\end{equation}
\textit{ where $\bm{K}_{2}$ is given by \eqref{eq:39eq}.\\
Under \eqref{eq:48eq}, the matrix $\bm{A}^{-1}(z)$ has the Laurent expansion \eqref{eq:29eq} about  $z=1$ with $m=2$ and}
\begin{equation}\label{eq:49eq}
\bm{N}_{-2}=\bm{C}_{0\bot}\bm{C}_{1\bot}(\bm{B}_{1\bot}^{'}\bm{B}_{0\bot}^{'}\bm{A}^{[2]}\bm{C}_{0\bot}\bm{C}_{1\bot})^{-1}
\end{equation}
\textit{as leading matrix of the principal part}.\\ 
\textit{The matrix function $\bm{P}_2\bm{A}^{-1}(z)$ is analytic at $z=1$ for}
\begin{equation}\label{eq:50eq}
\bm{P}_2=-\begin{bmatrix}
\bm{0}, &\bm{0}, &\bm{I}
\end{bmatrix}
\begin{bmatrix}
2\bm{P}_1, &\bm{0}, &\bm{I}\\
\bm{0}, &2\bm{\Pi}_2, &\bm{I}\\
\bm{I}, &\bm{I}, &\bm{0}
\end{bmatrix}
\begin{bmatrix}
\bm{0}\\\bm{0}\\\bm{I}
\end{bmatrix}
\end{equation}
\textit{where}
\begin{equation}\label{eq:51eq}
\bm{\Pi}_2=(\bm{A}^{+}\bm{A}^{(1)}\bm{C}_{0\bot}\bm{C}_{1\bot})^{\top}
\end{equation}
\textit{The  rank of $\bm{P}_2$ is}
\begin{equation}\label{eq:52eq}
r(\bm{P}_{2})=r(\bm{C}_{0})-r(\bm{A}^{+}\bm{A}^{(1)}\bm{C}_{0\bot}\bm{C}_{1\bot})
\end{equation}
\item \textit{The unit root $z=1$ is a  triple pole of $\bm{A}^{-1}(z)$, that is $m=3$, if}  
\begin{equation}\label{eq:53eq}
det\bm{K}_{2}=0
\end{equation}
\begin{equation}\label{eq:54eq}
det\bm{K}_{3}\neq 0
\end{equation}
\textit{where $\bm{K}_{3}=(\bm{B}_{2\bot}^{'}\bm{B}_{1\bot}^{'}\bm{B}_{0\bot}^{'}\bm{A}^{[3]}\bm{C}_{0\bot}\bm{C}_{1\bot}\bm{C}_{2\bot})$.\\
Under \eqref{eq:54eq}, the matrix $\bm{A}^{-1}(z)$  has the Laurent expansion \eqref{eq:29eq} about $z=1$ with $m=3$ and}
\begin{equation}\label{eq:55eq}
\bm{N}_{-3}=\bm{C}_{0\bot}\bm{C}_{1\bot}\bm{C}_{2\bot}(\bm{B}_{2\bot}^{'}\bm{B}_{1\bot}^{'}\bm{B}_{0\bot}^{'}\bm{A}^{[3]}\bm{C}_{0\bot}\bm{C}_{1\bot}\bm{C}_{2\bot})^{-1}\bm{B}_{2\bot}^{'}\bm{B}_{1\bot}^{'}\bm{B}_{0\bot}^{'}
\end{equation}
\textit{as leading matrix of the principal part.\\
The matrix function $\bm{P}_3\bm{A}^{-1}(z)$ is analytic at $z=1$ for}
\begin{equation}\label{eq:56eq}
\bm{P}_3=-\begin{bmatrix}
\bm{0}, &\bm{0}, &\bm{I}
\end{bmatrix}
\begin{bmatrix}
2\bm{P}_1, &\bm{0}, &\bm{I}\\
\bm{0}, &2\bm{\Pi}_3, &\bm{I}\\
\bm{I}, &\bm{I}, &\bm{0}
\end{bmatrix}
\begin{bmatrix}
\bm{0}\\\bm{0}\\\bm{I}
\end{bmatrix}
\end{equation}
\textit{where}
\begin{equation}\label{eq:57eq}
\bm{\Pi}_3=2[(\bm{A}^{+}\bm{A}^{(1)}\bm{C}_{0\bot})^{\top}:(\bm{A}^{+}\bm{A}^{[2]}\bm{C}_{0\bot}\bm{C}_{1\bot}\bm{C}_{2\bot})^{\top}]
\end{equation}
\textit{The rank of $\bm{P}_3$ is }
\begin{equation}\label{eq:58eq}
r(\bm{P}_3)=r(\bm{C}_{0})-r([\bm{A}^{+}\bm{A}^{(1)}\bm{C}_{0\bot}, \enspace\bm{A}^{+}\bm{A}^{[2]}\bm{C}_{0\bot}\bm{C}_{1\bot}\bm{C}_{2\bot}])
\end{equation}
\item \textit{The unit root $z=1$ is a  fourth order pole of $\bm{A}^{-1}(z)$, that is $m=4$, if }
\begin{equation}\label{eq:59eq}
det\bm{K}_{3}=0
\end{equation}
\begin{equation}\label{eq:60eq}
det\bm{K}_{4}\neq 0
\end{equation}
\textit{where $\bm{K}_{4}=(\bm{B}_{3\bot}^{'}\bm{B}_{2\bot}^{'}\bm{B}_{1\bot}^{'}\bm{B}_{0\bot}^{'}\bm{A}^{[4]}\bm{C}_{0\bot}\bm{C}_{1\bot}\bm{C}_{2\bot}\bm{C}_{3\bot})$.\\
Under \eqref{eq:60eq}, the matrix $\bm{A}^{-1}(z)$ has the Laurent expansion \eqref{eq:29eq} about $z=1$ with $m=4$ and} \begin{equation}\label{eq:61eq}
\bm{N}_{4}=\bm{C}_{0\bot}\bm{C}_{1\bot}\bm{C}_{2\bot}\bm{C}_{3\bot}(\bm{B}_{3\bot}^{'}\bm{B}_{2\bot}^{'}\bm{B}_{1\bot}^{'}\bm{B}_{0\bot}^{'}\bm{A}^{[4]}\bm{C}_{0\bot}\bm{C}_{1\bot}\bm{C}_{2\bot}\bm{C}_{3\bot})^{-1}\bm{B}_{3\bot}^{'}\bm{B}_{2\bot}^{'}\bm{B}_{1\bot}^{'}\bm{B}_{0\bot}^{'}
\end{equation}
\textit{as leading matrix of the principal part.\\
The matrix function $\bm{P}_4\bm{A}^{-1}(z)$ is analytic at $z=1$ for}
\begin{equation}\label{eq:62eq}
\bm{P}_4=-\begin{bmatrix}
\bm{0}, &\bm{0}, &\bm{I}
\end{bmatrix}
\begin{bmatrix}
2\bm{P}_1, &\bm{0}, &\bm{I}\\
\bm{0}, &2\bm{\Pi}_4, &\bm{I}\\
\bm{I}, &\bm{I}, &\bm{0}
\end{bmatrix}
\begin{bmatrix}
\bm{0}\\\bm{0}\\\bm{I}
\end{bmatrix}
\end{equation}
\textit{where}
\begin{equation}\label{eq:63eq}
\bm{\Pi}_4=2(\bm{\Pi}_{3,4}:(\bm{A}^{+}\bm{A}^{[3]}\bm{C}_{0\bot}\bm{C}_{1\bot}\bm{C}_{2\bot}\bm{C}_{3\bot})^{\top})
\end{equation}
\textit{and}
\begin{equation}\label{eq:64eq}
\bm{\Pi}_{3,4}=2((\bm{A}^{+}\bm{A}^{(1)}\bm{C}_{0\bot})^{\top}:(\bm{A}^{+}\bm{A}^{[2]}\bm{C}_{0\bot})^{\top})
\end{equation}
\textit{The rank of $\bm{P}_4$ is}
\begin{equation}\label{eq:65eq}
r(\bm{P}_4)=r(\bm{C}_{0})-r([\bm{A}^{+}\bm{A}^{(1)}\bm{C}_{0\bot}
,\enspace \bm{A}^{+}\bm{A}^{[2]}\bm{C}_{0\bot}])+r(\bm{\Xi})-r([\bm{\Xi}, \bm{\Gamma}])    
\end{equation}
\end{enumerate}
\textit{with $\bm{\Gamma}\bm{\Gamma}^{+}=\bm{\Pi}_{3,4}$ and $\bm{\Xi}\bm{\Xi}^{+}=(\bm{A}^{[3]}\bm{C}_{0\bot}\bm{C}_{1\bot}\bm{C}_{2\bot}\bm{C}_{3\bot})^{\top}$}
\\
\textbf{Proof}\\
Let $m=1$, then the following 
\begin{equation}\label{eq:66eq}
\bm{N}_{-m+1}\bm{A}+\bm{N}_{-m}\bm{A}^{(1)}=\bm{I}
\end{equation}
holds true because of \eqref{eq:32eq}, and the other way around.\\
Pre and post-multiplication of \eqref{eq:66eq} by $\bm{C}_{0\bot}^{+}$  and $\bm{C}_{0\bot}$, respectively, and making use of \eqref{eq:34eq} leads to the equation
\begin{equation}\label{eq:67eq}
\bm{Z}_{m}\bm{K}_{1}=\bm{I}
\end{equation}
which is consistent if and only if $\bm{K}_{1}$ is non-singular. Solving for $\bm{Z}_{m}$ yields
\begin{equation}\label{eq:68eq}
\bm{Z}_{m}=\bm{K}_{1}^{-1}
\end{equation}
and \eqref{eq:45eq} follows from \eqref{eq:34eq}, accordingly.\\ 
About the simple pole located at $z=1$ the Laurent expansion \eqref{eq:29eq} takes the form
\begin{equation}\label{eq:69eq}
\bm{A}^{-1}(z)=(z-1)^{-1}\bm{N}_{-1}+\bm{M}(z)
\end{equation}
with $\bm{N}_{-1}$ given by \eqref{eq:45eq}. \\
By inspection of \eqref{eq:45eq} it is easy to see that 
\begin{equation}\label{eq:70eq}
\bm{P}_{1}\bm{N}_{-1}=\bm{0}
\end{equation}
It follows that $\bm{P}_{1}\bm{A}^{-1}(z)=\bm{P}_{1}\bm{M}(z)$ is analytic at $z=1$.\\
Turning back to \eqref{eq:67eq}, if $\bm{K}_{1}$ is singular then the equation becomes inconsistent and we are facing a multiple pole. It follows that the right-hand sides of both \eqref{eq:66eq} and  \eqref{eq:67eq} are no longer identity matrices, but null matrices instead. Eventually, the homogeneous equation
\begin{equation}\label{eq:71eq}
\bm{Z}_{m}\bm{B}_{1}=\bm{0}
\end{equation}
takes the place of \eqref{eq:67eq}. Equation \eqref{eq:71eq} pairs off with
\begin{equation}\label{eq:72eq}
\bm{C}_{1}^{'}\bm{Z}_{m}=\bm{0}
\end{equation}
which follows from (34) by using the same argument. Solving the systems \eqref{eq:71eq} and \eqref{eq:72eq} for $\bm{Z}_{m}$ yields
\begin{equation}\label{eq:33eq}
\bm{Z}_{m}=\bm{C}_{1\bot}\bm{C}_{1\bot}^{+}\bm{\Psi}_{m}(\bm{B}_{1\bot}^{'})^{+}\bm{B}_{1\bot}^{'}=\bm{C}_{1\bot}\bm{\Phi}_{m}\bm{B}_{1\bot}^{'}
\end{equation}
for some $\bm{\Psi}_{m}$, with $\bm{\Phi}_{m}=\bm{C}_{1\bot}^{+}\bm{\Psi}_{m}(\bm{B}_{1\bot}^{'})^{+}$. The representation
\begin{equation}\label{eq:74eq}
\bm{N}_{-m}=\bm{C}_{0\bot}\bm{C}_{1\bot}\bm{\Phi}_{m}\bm{B}_{1\bot}^{'}\bm{B}_{0\bot}^{'}
\end{equation}
follows from \eqref{eq:34eq}, accordingly.\\
Now, let $m=2$. Then, the following
\begin{equation}\label{eq:75eq}
\bm{N}_{-m+2}\bm{A}+\bm{N}_{-m+1}\bm{A}^{(1)}+\frac{1}{2}\bm{N}_{-m}\bm{A}^{(2)}=\bm{I}
\end{equation}
holds true because of \eqref{eq:32eq}, and the other way around.\\
Pre and post-multiplying \eqref{eq:75eq} by $\bm{C}_{1\bot}^{+}\bm{C}_{0\bot}^{+} $ and $\bm{C}_{1\bot}\bm{C}_{0\bot}$, respectively, and making use of \eqref{eq:148eq} and \eqref{eq:74eq}, leads to the equation  
\begin{equation}\label{eq:76eq}
\bm{\Phi}_{m}\bm{K}_{2}=\bm{I}
\end{equation}
which is consistent if and only if $\bm{K}_{2}$ is non-singular. Solving for $\bm{\Phi}_{m}$ yields
\begin{equation}\label{eq:77eq}
\bm{\Phi}_{m}=\bm{K}_{2}^{-1}
\end{equation}
and \eqref{eq:49eq} follows from \eqref{eq:74eq}, accordingly.  \\
About the double pole located at $z=1$ the Laurent expansion \eqref{eq:29eq} takes the form
\begin{equation}\label{eq:78eq}
\bm{A}^{-1}(z)=(z-1)^{-2}\bm{N}_{-2}(z-1)^{-1}\bm{N}_{-1}+\bm{M}(z)
\end{equation}
with $\bm{N}_{-2}$ given by \eqref{eq:49eq}. Then, taking into account \eqref{eq:148eq} and \eqref{eq:49eq}, it is easy to see that 
\begin{equation}\label{eq:79eq}
\bm{P}_{1}[(z-1)^{-2}\bm{N}_{-2}(z-1)^{-1}(\bm{N}_{-1}+\bm{A}^{+}\bm{A}^{(1)}\bm{N}_{-2})]=\bm{0}
\end{equation}
Upon noting that
\begin{equation}\label{eq:80eq}
\bm{\Pi}_{2}(\bm{A}^{+}\bm{A}^{(1)}\bm{N}_{-2})=\bm{0}
\end{equation}
where $\bm{\Pi}_{2}$ is given by \eqref{eq:51eq}, applying Lemma A.1 in Appendix A to $\bm{P}_{1}$ and $\bm{\Pi}_{2}$ yields a projector $\bm{P}_{2}=2(\bm{P}_{1}:\bm{\Pi}_{2})$ such that $\bm{P}_{2}\bm{A}^{-1}(z)=\bm{P}_{2}\bm{M}(z)$ is analytic at $z=1$.\\
The expression \eqref{eq:50eq} follows from \eqref{eq:137eq} in Appendix A.\\
As for the rank of $\bm{P}_{2}$, formula \eqref{eq:1391eq}  applies yielding
\begin{equation}
r(\bm{P}_{2})=r(\bm{C}_{0})+r(\bm{\Pi}_{2})-n= r(\bm{C}_{0})-r(\bm{A}^{+}\bm{A}^{(1)}\bm{C}_{0\bot}\bm{C}_{1\bot})
\end{equation}
as 
\begin{equation*}
\bm{\Pi}_{2}\bm{P}_{1}^{\top}=\bm{\Theta}_{1}^{\top}\bm{P}_{1}^{\top}=(\bm{I}-\bm{\Theta}_{1}\bm{\Theta}_{1}^{+})\bm{P}_{1}^{\top}=(\bm{I}-(\bm{\Theta}_{1})^{+'}\bm{\Theta}_{1}')\bm{P}_{1}^{\top}
\end{equation*}
\begin{equation}\label{eq:case2}
=\bm{P}_{1}^{\top}-(\bm{\Theta}_{1})^{+'}\bm{C}_{1\bot}'\bm{C}_{0\bot}'
(\bm{A}^{(1)})'(\bm{A}^{+})'\bm{P}_{1}^{\top}=\bm{P}_{1}^{\top}
\end{equation}
where $\bm{\Theta}_{1}=\bm{A}^{+}\bm{A}^{(1)}\bm{C}_{0\bot}\bm{C}_{1\bot}$, $\bm{P}_{1}^{\top}=\bm{C}_{0\bot}\bm{C}_{0\bot}^{+}$, $(\bm{A}^{+})'=(\bm{B}_{0}^{+})'(\bm{C}_{0}'\bm{C}_{0})^{-1}\bm{C}_{0}'$. Since
$(\bm{A}^{+})'\bm{P}_{1}^{\top}=\bm{0}$, the result  $\bm{\Theta}_{1}'\bm{P}_{1}^{\top}=\bm{0}$ follows as a by-product.\\
Turning back to \eqref{eq:76eq}, if $\bm{K}_{2}$ is singular, then the equation becomes inconsistent and we are facing  a pole of order higher than two.  It follows that the right-hand sides of \eqref{eq:75eq} and \eqref{eq:76eq} are no longer identity matrices but null matrices instead. Eventually, the homogeneous equation 
\begin{equation}\label{eq:82eq}
\bm{\Phi}_{m}\bm{B}_{2}=\bm{0}
\end{equation}
takes the place of \eqref{eq:76eq}. Equation \eqref{eq:82eq} pairs off with
\begin{equation}\label{eq:83eq}
\bm{C}_{2}^{'}\bm{\Phi}_{m}=\bm{0}
\end{equation}
which follows from (34) by using the same argument. Solving the systems \eqref{eq:82eq} and \eqref{eq:83eq} for $\bm{\Phi}_{m}$ yields 
\begin{equation}\label{eq:84eq}
\bm{\Phi}_{m}=\bm{C}_{2\bot}\bm{C}_{2\bot}^{+}\bm{\Psi}_{m}(\bm{B}_{2\bot}^{'})^{+}\bm{B}_{2\bot}^{'}=\bm{C}_{2\bot}\bm{Z}_{m}\bm{B}_{2\bot}^{'}
\end{equation}
for some $\bm{\Psi}_{m}$, with $\bm{Z}_{m}=\bm{C}_{2\bot}^{+}\bm{\Psi}_{m}(\bm{B}_{2\bot}^{'})^{+}$ .  The representation
\begin{equation}\label{eq:85eq}
\bm{N}_{-m}=\bm{C}_{0\bot}\bm{C}_{1\bot}\bm{C}_{2\bot}\bm{Z}_{m}\bm{B}_{2\bot}^{'}\bm{B}_{1\bot}^{'}\bm{B}_{0\bot}^{'}
\end{equation}
follows from \eqref{eq:74eq}, accordingly.\\
Now, let $m=3$. Then, the following 
\begin{equation}\label{eq:86eq}
\bm{N}_{-m+3}\bm{A}+\bm{N}_{-m+2}\bm{A}^{(1)}+\frac{1}{2}\bm{N}_{-m+1}\bm{A}^{(2)}+\frac{1}{6}\bm{N}_{-m}\bm{A}^{(3)}=\bm{I}
\end{equation}
holds true because of \eqref{eq:32eq} and the other way around. \\
Pre and post-multiplying \eqref{eq:86eq} by $\bm{F}_{1}^{-}=\bm{C}_{2\bot}^{+}\bm{C}_{1\bot}^{+}\bm{C}_{0\bot}^{+}$ and $\bm{F}_{1}=\bm{C}_{0\bot}\bm{C}_{1\bot}\bm{C}_{2\bot}$, respectively, yields
\begin{equation}\label{eq:87eq}
\bm{F}_{1}^{-}\bm{N}_{-m+2}\bm{A}^{(1)}\bm{F}_{1}+\frac{1}{2}\bm{F}_{1}^{-}\bm{N}_{-m+1}\bm{A}^{(2)}\bm{F}_{1}+\frac{1}{6}\bm{F}_{1}^{-}\bm{N}_{-m}\bm{A}^{(3)}=\bm{I}
\end{equation}
as $\bm{A}\bm{F}_{1}=\bm{0}$. Now, replacing $\bm{N}_{-m+2}$, given by \eqref{eq:152eq}, into \eqref{eq:87eq} gives
\begin{equation}\label{eq:88eq}
\bm{F}_{1}^{-}\bm{N}_{-m+1}\bm{A}^{[2]}\bm{F}_{1}-\frac{1}{2}\bm{F}_{1}^{-}\bm{N}_{-m}\bm{A}^{(2)}\bm{A}^{+}\bm{A}^{(1)}+ \frac{1}{6}\bm{F}_{1}^{-}\bm{N}_{-m}\bm{A}^{(3)}\bm{F}_{1} =\bm{I}
\end{equation}
as $\bm{F}_{1}^{-}\bm{A}^{+}$ and  $\bm{B}_{0\bot}^{'}\bm{A}^{(1)}\bm{F}_{1}$ are null matrices.\\
Then, replacing $\bm{N}_{-m+1}$, given by \eqref{eq:148eqzl}, into \eqref{eq:88eq} gives
\begin{equation}\label{eq:89eq}
\bm{F}_{1}^{-}\bm{N}_{-m}\bm{A}^{[3]}\bm{F}_{1} =\bm{I}
\end{equation}
as $\bm{F}_{1}^{-}\bm{\Theta}_{1}$ and $\bm{B}_{1\bot}^{'}\bm{B}_{0\bot}^{'}\bm{A}^{[2]}\bm{F}_{1}$ are null matrices. \\
Equation \eqref{eq:89eq}, in light of \eqref{eq:85eq} and \eqref{eq:4eq}, can be also written as follows
\begin{equation}\label{eq:90eq}
\bm{Z}_{m}\bm{K}_{3}=\bm{I}
\end{equation}
as $\bm{F}_{1}^{-}\bm{N}_{-m}=\bm{Z}_{m}\bm{B}_{2\bot}^{'}\bm{B}_{1\bot}^{'}\bm{B}_{0\bot}^{'}$.\\
Equation \eqref{eq:90eq} is consistent if and only if $\bm{K}_{3}$ is non-singular. Solving \eqref{eq:90eq} yields
\begin{equation}\label{eq:91eq}
\bm{Z}_{m}=\bm{K}_{3}^{-1}
\end{equation}
and \eqref{eq:55eq} follows, accordingly.\\
About the 3-rd order pole located at $z=1$, the Laurent expansion  \eqref{eq:29eq} takes the form
\begin{equation}\label{eq:92eq}
\bm{A}^{-1}(z)=(z-1)^{-3}\bm{N}_{-3}+(z-1)^{-2}\bm{N}_{-2}+(z-1)^{-1}\bm{N}_{-1}+\bm{M}(z)
\end{equation}
with $\bm{N}_{-3}$ given by \eqref{eq:55eq}. Then, taking into account formulas \eqref{eq:148eq}, \eqref{eq:41eq} and  \eqref{eq:55eq}, it is easy to verify that 
\begin{equation}\label{eq:93eq}
\bm{P}_{1}[(z-1)^{-3}\bm{N}_{-3}+(z-1)^{-2}(\bm{N}_{-2}+\bm{A}^{+}\bm{A}^{(1)}\bm{N}_{3})+(z-1)^{-1}(\bm{N}_{-1}+\bm{A}^{+}\bm{A}^{[2]}\bm{N}_{-3}+\bm{\Xi}_{1})]=\bm{0}
\end{equation}
where $\bm{\Xi}_{1}=-\bm{A}^{+}\bm{A}^{(1)}\bm{N}_{-3}\bm{A}^{(1)}\bm{A}^{+}+\bm{A}^{+}\bm{A}^{(1)}\bm{C}_{0\bot}\bm{S}_{1,1}\bm{B}_{0\bot}^{'}$.\\
Upon noting that
\begin{equation}\label{eq:94eq}
(\bm{A}^{+}\bm{A}^{[2]}\bm{C}_{0\bot}\bm{C}_{1\bot}\bm{C}_{2\bot})^{\top}\bm{A}^{+}\bm{A}^{[2]}\bm{N}_{-3}=\bm{0}
\end{equation}
\begin{equation}\label{eq:95eq}
(\bm{A}^{+}\bm{A}^{(1)}\bm{C}_{0\bot})^{\top}((z-1)^{-2}\bm{A}^{+}\bm{A}^{(1)}\bm{N}_{-3}+(z-1)^{-1}\bm{\Xi}_{1})=\bm{0}
\end{equation}
the application of Lemma A.1 in Appendix A leads to the conclusion
\begin{equation}\label{eq:96eq}
\bm{\Pi}_{3}[(z-1)^{-2}\bm{A}^{+}\bm{A}^{(1)})\bm{N}_{-3}+(z-1)^{-1}(\bm{A}^{+}\bm{A}^{[2]}\bm{N}_{-3}+\bm{\Xi}_{1})]=\bm{0}
\end{equation}
where $\bm{\Pi}_{3}$ is the matrix given by \eqref{eq:57eq}, and eventually that
\begin{equation}\label{eq:97eq}
\bm{P}_{3}\bm{A}^{-1}(z)=\bm{P}_{(3)})\bm{M}(z)
\end{equation}
where $\bm{P}_{3}=2(\bm{P}_{1}:\bm{\Pi}_{3})$. In light of \eqref{eq:97eq}, $\bm{P}_{3}\bm{A}^{-1}(z)$ is analytic at $z=1$. The expression  \eqref{eq:56eq} follows from \eqref{eq:137eq} in Appendix A.\\ The rank of $\bm{P}_{3}$ can be established following an argument similar to that used to obtain the rank of $\bm{P}_{2}$. Applying formula \eqref{eq:1391eq} in Appendix A yields
\begin{equation}\label{eq:case3}
r(\bm{P}_{3})=r(\bm{P}_{1})+r(\bm{\Pi}_{3})-n
\end{equation}
as 
\begin{equation*}
\bm{\Pi}_{3}\bm{P}_{1}^{\top}=2(\widetilde{\bm{\Theta}}_{1}^{\top}:\bm{\Theta}_{2}^{\top})\bm{P}_{1}^{\top}=\widetilde{\bm{\Theta}}_{1}^{\top}\bm{P}_{1}^{\top}-(\widetilde{\bm{\Theta}}_{1}^{\top}-\bm{\Theta}_{2}^{\top}\widetilde{\bm{\Theta}}_{1}^{\top})^{+}(\widetilde{\bm{\Theta}}_{1}^{\top}-\bm{\Theta}_{2}^{\top}\bm{\widetilde{\Theta}}_{1}^{\top})^{+}\bm{P}_{1}^{\top}
\end{equation*}
\begin{equation}\label{eq:cc3}
=\bm{P}_{1}^{\top}-(\widetilde{\bm{\Theta}}_{1}^{\top}-\bm{\Theta}_{2}^{\top}\widetilde{\bm{\Theta}}_{1}^{\top})^{+}(\bm{P}_{1}^{\top}-\bm{\Theta}_{2}^{\top}\bm{P}_{1}^{\top})=\bm{P}_{1}^{\top}
\end{equation}
Here $\widetilde{\bm{\Theta}}_{1}^{\top}=\bm{A}^{+}\bm{A}^{(1)}\bm{C}_{0\bot}$,  $\bm{\Theta}_{2}=\bm{A}^{+}\bm{A}^{[2]}\bm{C}_{0\bot}\bm{C}_{1\bot}\bm{C}_{2\bot}$ and use has been made of \eqref{eq:138seq1} in Appendix A, of \eqref{eq:case2} above and of the equality $\bm{\Theta}_{2}^{\top}\bm{P}_{1}^{\top}=\bm{P}_{1}^{\top}$ which can be proved by using the same approach followed to obtain \eqref{eq:case2}. Thanks to \eqref{eq:fg}, formula \eqref{eq:case3} can be worked out as follows
\begin{equation}
r(\bm{P}_{3})=r(\bm{C}_{0})+r(\widetilde{\bm{\Theta}}_{1}^{\top}:\bm{\Theta}_{2}^{\top})-n=r(\bm{C}_{0})-r([\widetilde{\bm{\Theta}}_{1}^{\top}, \bm{\Theta}_{2}])
\end{equation}
Turning back to \eqref{eq:90eq}, if $\bm{K}_{3}$ is singular then the equation becomes inconsistent and we are facing a pole of order higher than three. It follows that the right-hand sides of
\eqref{eq:86eq} and \eqref{eq:90eq} are no longer identity matrices but null matrices instead. Eventually, the homogeneous equation 
\begin{equation}\label{eq:100eq}
\bm{Z}_{m}\bm{B}_{3}=\bm{0}
\end{equation}
takes the place of the former \eqref{eq:90eq}. Equation \eqref{eq:100eq} pairs off with 
\begin{equation}\label{eq:101eq}
\bm{C}_{3\bot}^{'}\bm{Z}_{m}=\bm{0}
\end{equation}
which follows from (34) making use of the same argument. Solving \eqref{eq:100eq} and \eqref{eq:101eq} for $\bm{Z}_{m}$ yields 
\begin{equation}\label{eq:102eq}
\bm{Z}_{m}=\bm{C}_{3\bot}\bm{C}_{3\bot}^{+}\widetilde{\bm{\Psi}}_{m}(\bm{B}_{3\bot}^{'})^{+}\bm{B}_{3\bot}^{'}=\bm{C}_{3\bot}\widetilde{\bm{\Phi}}_{m}\bm{B}_{3\bot}^{'}
\end{equation}
for some $\widetilde{\bm{\Psi}}_{m}$, with $\widetilde{\bm{\Phi}}_{m}=\bm{C}_{3\bot}^{+}\widetilde{\bm{\Psi}}_{m}(\bm{B}_{3\bot}^{'})^{+}$. The representation 
\begin{equation}\label{eq:103eq}
\bm{N}_{-m}=\bm{C}_{0\bot}\bm{C}_{1\bot}\bm{C}_{2\bot}\bm{C}_{3\bot}\widetilde{\bm{\Phi}}_{m}\bm{B}_{3\bot}^{'}\bm{B}_{2\bot}^{'}\bm{B}_{1\bot}^{'}\bm{B}_{0\bot}^{'}
\end{equation}
follows from \eqref{eq:85eq}, accordingly. \\
Now, let $m=4$. Then, the following
\begin{equation}\label{eq:104eq}
\bm{N}_{-m+4}\bm{A}+\bm{N}_{-m+3}\bm{A}^{(1)}+\frac{1}{2}\bm{N}_{-m+2}\bm{A}^{(2)}+\frac{1}{6}\bm{N}_{-m+1}\bm{A}^{(3)}+\frac{1}{24}\bm{N}_{-m}\bm{A}^{(4)}=\bm{I}
\end{equation}
holds true because of \eqref{eq:32eq} and the other way around. \\
Pre and post-multiplying \eqref{eq:104eq} by $\bm{F}_{2}^{-}=\bm{C}_{3\bot}^{+}\bm{C}_{2\bot}^{+}\bm{C}_{1\bot}^{+}\bm{C}_{0\bot}^{+}$ and $\bm{F}_{2}=\bm{C}_{3\bot}\bm{C}_{2\bot}\bm{C}_{1\bot}\bm{C}_{0\bot}$, respectively, yields
\begin{equation}\label{eq:105eq}
\bm{F}_{2}^{-}\bm{N}_{-m+3}\bm{A}^{(1)}\bm{F}_{2}+\frac{1}{2}\bm{F}_{2}^{-}\bm{N}_{-m+2}\bm{A}^{(2)}\bm{F}_{2}+\frac{1}{6}\bm{F}_{2}^{-}\bm{N}_{-m+1}\bm{A}^{(3)}\bm{F}_{2}+\frac{1}{24}\bm{F}_{2}^{-}\bm{N}_{-m}\bm{A}^{(4)}\bm{F}_{2}=\bm{I}
\end{equation}
as $\bm{A}\bm{F}_{2}=\bm{0}$. \\
By replacing $\bm{N}_{-m+3}$, given by \eqref{eq:159eq} into \eqref{eq:105eq} gives 
\begin{equation*}
\bm{F}_{2}^{-}\bm{N}_{-m+2}\bm{A}^{[2]}\bm{F}_{2}-\frac{1}{2}\bm{F}_{2}^{-}\bm{N}_{-m+1}\bm{A}^{(2)}\bm{A}^{+}\bm{A}^{(1)}\bm{F}_{2}-\frac{1}{6}\bm{F}_{2}^{-}\bm{N}_{-m}\bm{A}^{(3)}\bm{A}^{+}\bm{A}^{(1)}\bm{F}_{2}+
\end{equation*}
\begin{equation}\label{eq:106eq}
\frac{1}{6}\bm{F}_{2}^{-}\bm{N}_{-m+1}\bm{A}^{(3)}\bm{F}_{2}+\frac{1}{24}\bm{F}_{2}^{-}\bm{N}_{-m}\bm{A}^{(4)}\bm{F}_{2}=\bm{I}
\end{equation}
as $\bm{F}_{2}^{-}\bm{A}^{+}$ and $\bm{B}_{0\bot}^{'}\bm{A}^{(1)}\bm{F}_{2}$ are null matrices.\\
Then, replacing $\bm{N}_{-m+2}$, given by \eqref{eq:ttz}, into \eqref{eq:106eq} gives
\begin{equation*}
\bm{F}_{2}^{-}\bm{N}_{-m+1}\bm{A}^{[3]}\bm{F}_{2}-\bm{F}_{2}^{-}\bm{N}_{-m}\bm{A}^{(1)}\bm{A}^{+}\bm{A}^{(1)}\bm{A}^{[2]}\bm{F}_{2}-\bm{F}_{2}^{-}\bm{N}_{-m}\bm{A}^{[3]}\bm{\Theta}_{1}\bm{A}^{[2]}\bm{F}_{2}+
\end{equation*}
\begin{equation*}
-\bm{F}_{2}^{-}\bm{N}_{-m}\bm{A}^{[2]}\bm{A}^{+}\bm{A}^{[2]}\bm{F}_{2} -\bm{F}_{2}^{-}\bm{N}_{-m}\bm{A}^{(1)}\bm{A}^{+}\bm{A}^{[2]}\bm{\Theta}_{1}\bm{A}^{[2]}\bm{F}_{2}-\bm{F}_{2}^{-}\bm{N}_{-m}\bm{A}^{[2]}\bm{\Theta}_{1}\bm{A}^{[2]}\bm{\Theta}_{1}\bm{A}^{[2]}\bm{F}_{2}
\end{equation*}
\begin{equation}\label{eq:107eq}
-\frac{1}{6}\bm{F}_{2}^{-}\bm{N}_{-m}\bm{A}^{(3)}\bm{A}^{+}\bm{A}^{(1)}\bm{F}_{2}+\frac{1}{24}\bm{F}_{2}^{-}\bm{N}_{-m}\bm{A}^{(4)}\bm{F}_{2}
\end{equation}
as  $\bm{F}_{2}^{-}\bm{\Theta}_{1}$ and $\bm{B}_{1\bot}^{'}\bm{B}_{0\bot}^{'}\bm{A}^{[2]}\bm{F}_{2}$ are null matrices.\\
Finally, replacing $\bm{N}_{-m+1}$, given by \eqref{eq:148eqzt}, into \eqref{eq:107eq} gives
\begin{equation}\label{eq:fine}
\bm{F}_{2}^{-}\bm{N}_{-m}\bm{A}^{[4]}\bm{F}_{2}=\bm{I}
\end{equation}
Equation \eqref{eq:fine}, taking into account   \eqref{eq:103eq} and \eqref{eq:4eq}, can be also written as
\begin{equation}\label{eq:109eq}
\widetilde{\bm{\Phi}}_{m}\bm{K}_{4}=\bm{I}
\end{equation}
because $\bm{F}_{2}^{-}\bm{N}_{m}=\widetilde{\bm{\Phi}}_{m}\bm{B}_{3\bot}^{'}\bm{B}_{2\bot}^{'} \bm{B}_{1\bot}^{'}\bm{B}_{0\bot}^{'}$.\\
Equation \eqref{eq:109eq} is consistent if and only if $\bm{K}_{4}$ is non-singular. Solving \eqref{eq:109eq} yields
\begin{equation}\label{eq:110eq}
\widetilde{\bm{\Phi}}_{m}=\bm{K}_{4}^{-1}
\end{equation}
and \eqref{eq:61eq} follows, accordingly. \\
About the 4-th order pole located at $z=1$, the Laurent expansion  \eqref{eq:29eq} takes the form
\begin{equation}\label{eq:110seq}
\bm{A}^{-1}(z)=(z-1)^{-4}\bm{N}_{-4}+(z-1)^{-3}\bm{N}_{-3}+(z-1)^{-2}\bm{N}_{-2}+(z-1)^{-1}\bm{N}_{-1}+\bm{M}(z)
\end{equation}
with $\bm{N}_{-4}$ given by \eqref{eq:61eq}.  \\ Taking into account \eqref{eq:148eq}, \eqref{eq:41eq}, \eqref{eq:43seq} and \eqref{eq:61eq}, it is easy to see that 
\begin{center}
$\bm{P}_{1}[(z-1)^{-4}\bm{N}_{-4}+(z-1)^{-3}(\bm{N}_{-3}+\bm{A}^{+}\bm{A}^{(1)}\bm{N}_{-4})+(z-1)^{-2}(\bm{N}_{-2}+\bm{A}^{+}\bm{A}^{[2]}\bm{N}_{-4}+\widetilde{\bm{\Xi}}_{1})+$    
\end{center}
\begin{equation}\label{eq:111eq}
 +(z-1)^{-1}(\bm{N}_{-1}+\bm{A}^{+}\bm{A}^{[3]}\bm{N}_{-4}+\bm{\Xi}_{2}+\bm{\Xi}_{3})]=\bm{0}
\end{equation}
where
\begin{equation}\label{eq:112eq}
\widetilde{\bm{\Xi}}_{1}=-\bm{A}^{+}\bm{A}^{(1)}\bm{N}_{-4}\bm{A}^{(1)}\bm{A}^{+}+\bm{A}^{+}\bm{A}^{(1)}\bm{C}_{0\bot}\bm{S}_{1,1}\bm{B}_{0\bot}^{'},\end{equation}
\begin{equation}\label{eq:113eq}
\bm{\Xi}_{2}=-\bm{A}^{+}\bm{A}^{[2]}\bm{N}_{-4}\bm{A}^{(1)}\bm{A}^{+}+\bm{A}^{+}\bm{A}^{[2]}\bm{\Theta}_{1}\bm{A}^{[2]}\bm{N}_{-4}+\bm{A}^{+}\bm{A}^{[2]}\bm{C}_{0\bot}\bm{S}_{1,1}\bm{B}_{0\bot}^{'},\end{equation}
\begin{equation}\label{eq:114eq}
\bm{\Xi}_{3}=-\bm{A}^{+}\bm{A}^{(1)}\bm{N}_{-4}\bm{A}^{[2]}\bm{A}^{+}-\bm{A}^{+}\bm{A}^{(1)}\bm{C}_{0\bot}\bm{S}_{1,1}\bm{B}_{0\bot}^{'}\bm{A}^{(1)}\bm{A}^{+}+\bm{A}^{+}\bm{A}^{(1)}\bm{C}_{0\bot}\bm{S}_{2,1}\bm{B}_{0\bot}^{'},\end{equation}
Upon noting that
\begin{equation}\label{eq:115eq}
(\bm{A}^{+}\bm{A}^{[3]}\bm{C}_{0\bot}\bm{C}_{1\bot}\bm{C}_{2\bot}\bm{C}_{3\bot})^{\top}\bm{A}^{+}\bm{A}^{[3]}\bm{N}_{-4}=\bm{0}\end{equation}
\begin{equation}\label{eq:116eq}
(\bm{A}^{+}\bm{A}^{[2]}\bm{C}_{0\bot})^{\top}((z-1)^{-2}\bm{A}^{+}\bm{A}^{[2]}\bm{N}_{-4}+(z-1)^{-1}\bm{\Xi}_{2})=\bm{0}\end{equation}
\begin{equation}\label{eq:117eq}
(\bm{A}^{+}\bm{A}^{(1)}\bm{C}_{0\bot})^{\top}((z-1)^{-3}\bm{A}^{+}\bm{A}^{(1)}\bm{N}_{-4}+(z-1)^{-2}\widetilde{\bm{\Xi}}_{1}+(z-1)^{-1}\bm{\Xi}_{3})=\bm{0}\end{equation}
it follows from Lemma A1 in Appendix A that 
\begin{equation}\label{eq:118eq}
\bm{\Pi}_{3,4}((z-1)^{-3}\bm{A}^{+}\bm{A}^{(1)}\bm{N}_{-4}+(z-1)^{-2}(\bm{A}^{+}\bm{A}^{[2]}\bm{C}_{0\bot}+\widetilde{\bm{\Xi}}_{1})+(z-1)^{-1}(\bm{\Xi}_{2}+\bm{\Xi}_{3}))=\bm{0}\end{equation}
where $\bm{\Pi}_{3,4}$ is the matrix given by \eqref{eq:64eq}, and 
\begin{equation}\label{eq:119eq}
\bm{\Pi}_{4}[((z-1)^{-3}\bm{A}^{+}\bm{A}^{(1)}\bm{N}_{-4}+(z-1)^{-2}(\bm{A}^{+}\bm{A}^{[2]}\bm{N}_{-4}+\widetilde{\bm{\Xi}}_{1})+(z-1)^{-1}(\bm{A}^{+}\bm{A}^{[3]}\bm{N}_{-4}+\bm{\Xi}_{2}+\bm{\Xi}_{3})]=\bm{0}\end{equation}
where $\bm{\Pi}_{4}$ is the matrix given by \eqref{eq:63eq}.  Eventually, it follows that  
\begin{equation}\label{eq:120eq}
\bm{P}_{4}\bm{A}^{-1}(z)=\bm{P}_{4}\bm{M}(z)
\end{equation}
where $\bm{P}_{4}=2(\bm{P}_{1}:\bm{\Pi}_{4})$. In light of \eqref{eq:120eq}, $\bm{P}_{4}\bm{A}^{-1}(z)$ is analytic at $z=1$. The expression \eqref{eq:62eq} follows from \eqref{eq:137eq} in Appendix A. \\
The rank of $\bm{P}_{4}$ can be established following an argument similar to that used for $\bm{P}_{3}$. Applying formula \eqref{eq:1391eq}  in Appendix A yields
\begin{equation}\label{eq:120seq}
r(\bm{P}_{4})=r(\bm{P}_{1}:\bm{\Pi}_{4})=r(\bm{P}_{1})+r(\bm{\Pi}_{4})-r(\bm{P}_{1}+\bm{\Pi}_{4})=r(\bm{C}_{0})+r(\bm{\Pi}_{4})-n
\end{equation}
as it can be proved that $\bm{\Pi}_{4}\bm{P}_{1}^{\top}=\bm{P}_{1}^{\top}$ by repeating the argument of formula \eqref{eq:cc3}. 
Applying  \eqref{eq:139eq}  to $\bm{\Pi}_{4}$ yields
\begin{equation*}
r(\bm{\Pi}_{4})=r[\bm{\Pi}_{3,4}:(\bm{A}^{+}\bm{A}^{[3]}\bm{C}_{0\bot}\bm{C}_{1\bot}\bm{C}_{2\bot}\bm{C}_{3\bot})^{\top}]=r(\bm{\Pi}_{3,4})+
\end{equation*}
\begin{equation}\label{eq:122eq}
+r(\bm{A}^{+}\bm{A}^{[3]}\bm{C}_{0\bot}\bm{C}_{1\bot}\bm{C}_{2\bot}\bm{C}_{3\bot})^{\top}  
 -r(\bm{\Pi}_{3,4}+(\bm{A}^{+}\bm{A}^{[3]}\bm{C}_{0\bot}\bm{C}_{1\bot}\bm{C}_{2\bot}\bm{C}_{3\bot})^{\top})
\end{equation}
Here
\begin{equation}\label{eq:123eq}
r(\bm{\Pi}_{3,4})=r[(\bm{A}^{+}\bm{A}^{(1)}\bm{C}_{0\bot})^{\top}:(\bm{A}^{+}\bm{A}^{[2]}\bm{C}_{0\bot})^{\top}]=n-r[(\bm{A}^{+}\bm{A}^{(1)}\bm{C}_{0\bot}), \enspace (\bm{A}^{+}\bm{A}^{[2]}\bm{C}_{0\bot})]
\end{equation}
in light of \eqref{eq:fg} in Appendix A, and
\begin{equation}\label{eq:1233}
 r(\bm{\Pi}_{3,4}+(\bm{A}^{+}\bm{A}^{[3]}\bm{C}_{0\bot}\bm{C}_{1\bot}\bm{C}_{2\bot}\bm{C}_{3\bot})^{\top}]=r([\bm{\Gamma}, \bm{\Xi}])
\end{equation}
by setting $\bm{\Pi}_{3,4}=\bm{\Gamma}\bm{\Gamma}^{+}$ and $(\bm{A}^{+}\bm{A}^{[3]}\bm{C}_{0\bot}\bm{C}_{1\bot}\bm{C}_{2\bot}\bm{C}_{3\bot})^{\top}=\bm{\Xi}\bm{\Xi}^{+}$.\\
Thanks to \eqref{eq:122eq}, \eqref{eq:123eq} and \eqref{eq:1233}, formula \eqref{eq:120seq} can be worked out as follows
\begin{equation}
r(\bm{P}_{4})=r(\bm{C}_{0})-r[(\bm{A}^{+}\bm{A}^{(1)}\bm{C}_{0\bot}), \enspace (\bm{A}^{+}\bm{A}^{[2]}\bm{C}_{0\bot})]+r(\bm{\Xi})-r([\bm{\Gamma}, \bm{\Xi}])
\end{equation}
This proves \eqref{eq:65eq}. 
\begin{flushright}
$\square$
\end{flushright}

\section{The main theorem continued}{\label{sec:4sec}}
Thanks to the analytic toolkit we have settled in the previous two sections, we are eventually ready to give the following
\\ \textbf{Proof of Theorem 1.1}
\\The VAR Model in \eqref{eq:1eq} is a linear non-homogeneous difference-equation system in matrix form whose solution can be formally written as
\begin{equation}\label{eq:124eq}
\bm{y}_{t}=\bm{A}^{-1}(L)\bm{\varepsilon}_{t}+\bm{A}^{-1}(L)\bm{0}
\end{equation}
where the first term is a particular solution of the non-homogeneous equation and the second is the so called complementary solution. Both depend on the operator $\bm{A}^{-1}(L)$, and eventually on the matrix $\bm{A}^{-1}(z)$ as the algebras of the polynomial functions of the lag operator $L$ and of the complex variable $z$ are isomorphic (see, e.g.,~\cite{dhrymes1971distributed}). The particular solution $\bm{A}^{-1}(L)\bm{\varepsilon}_{t}$ is composed of  a (coloured) noise term and random walks up to the $m$-th order, where $m$ is the order of the pole of $\bm{A}^{-1}(z)$ at $z=1$, while the complementary solution is a polynomial in t of ($m$-1)-th degree: altogether they lead to an $m$-th order integrated process (see, e.g.,~\cite{faliva2008dynamic} Sections 1.8 and 2.3). 
As for the pole order ($m$), this is established by Theorem 3.2, and \eqref{eq:2seq} ensues accordingly. \\
As for \eqref{eq:4eq}-\eqref{eq:13eq}, the formulas are proved in Theorem 3.2 as well. \\
The cointegration relationship \eqref{eq:3eq} recovers stationary by annihilating the principal-part of $\bm{A}^{-1}(z)$. The matrices $\bm{P}_{m}$, for $m$=1, 2, 3, 4  as per \eqref{eq:14eq}-\eqref{eq:17eq}, are obtained in the aforesaid theorem, too. The cointegration ranks in formulas \eqref{eq:22eq}- \eqref{eq:424eq} are derived in the same theorem.  \begin{flushright}
$\square$
\end{flushright}
\section{Concluding Remarks}{\label{sec:5sec}}
The paper investigates unit-root VARs whose solutions are (co)-integrated processes up to 4-th order. This is in itself worthy of note when compared the extant literature which is almost entirely dedicated to first and second order processes. What is more, the algebraic apparatus set forth in the paper can be successfully applied to higher order processes by induction, thus paving the way to a wide-spread field of applications. It is worth noticing that the key issue of stationary recovering via cointegration is settled for integrated processes of increasing order via a cunning algebraic argument which hinges on the notion of parallel sum. \\
\section*{Appendix A}\label{sec:appA}
\singlespacing
\textbf{Parallel sum of matrices}
\\The notion of parallel sum plays a key role in the approach to cointegration devised in the paper. 
As the non-stationarity of the solution
\begin{equation}
y_{t}=\bm{A}^{-1}(L)\epsilon_{t}+\bm{A}^{-1}(L)0
\end{equation}
of a unit-root VAR model arises from the principal part $\sum_{j=-m}^{-1}\bm{N}_{j}(z-1)^{j}$ of the Laurent expansion of $\bm{A}^{-1}(z)$ through $\bm{A}^{-1}(L)$, a transformation $\bm{P}\bm{A}^{-1}(z)$ that annihilates the principal part recovers stationarity, that is
\begin{equation}
\bm{P}\bm{A}^{-1}(z)=\bm{P}\bm{M}(z)
\end{equation}
for some $\bm{P}$, then
\begin{equation}
\bm{P}\bm{A}^{-1}(L)=\bm{P}\bm{M}(L)\Rightarrow
y_{t}\sim I(0)
\end{equation}
In the case of a simple pole, the determination of an idempotent operator $\bm{P}=\bm{P}_{1}$ is straightforward (see Theorem 3.2).
In the case of a double pole (and more generally of a multiple pole) a stepwise procedure must been devised, that is \\
1 Step) check that $\bm{P}_{1}(\bm{N}_{2}, \bm{N}_{1}+\bm{\Phi})=\bm{0}$, where $\bm{\Phi}$
is a matrix that can be easily found.
\\
2 Step) determine an idempotent operator $\bm{\Pi}_{2}$ such that $\bm{\Pi}_{2}\bm{\Phi}=\bm{0}$.
\\
3 Step) Find out an idempotent operator $\bm{P}_{2}=f(\bm{P}_{1},\bm{\Pi}_{2})$ generated by a subset of vectors which are common to $\bm{P}_{1}$ and $\bm{\Pi}_{2}$, so that $\bm{P}_{2}(\bm{N}_{2}, \bm{N}_{1})=\bm{0}$.
The intended operator turns out to tally with twice the parallel sum of $\bm{P}_{1}$ and $\bm{\Pi}_{2}$, as the lemma which follows shows. 
The case of higher-order poles can be tackled by repeating the argument above as shown in Theorem 3.2.\\
The following  Lemma gives the results on parallel sums we are primarily interested in 
\\\textbf{Lemma A.1}\\
Let \textbf{V} and \textbf{W} be square matrices of the same order and \textbf{R}, \textbf{S} be two matrices satisfying
\begin{equation}\label{eq:125eq}
\bm{R}^{'}\bm{V}=\bm{0}
\end{equation}
\begin{equation}\label{eq:126eq}
\bm{S}^{'}\bm{W}=\bm{0}
\end{equation}
so that
\begin{equation}\label{eq:127eq}
\bm{P}_{R}=(\bm{R}^{'})^{+}\bm{R}^{'}
\end{equation}
\begin{equation}\label{eq:128eq}
\bm{P}_{S}=(\bm{S}^{'})^{+}\bm{S}^{'}
\end{equation}
are projection operators of $\bm{V}$ on $\bm{0}$ and of $\bm{W}$ on $\bm{0}$, respectively.\\
Then, the parallel sum $\bm{P}_{R}:\bm{P}_{S}$ of $\bm{P}_{R}$ and $\bm{P}_{S}$ is defined as 
\begin{equation}\label{eq:129eq}
\bm{P}_{R}:\bm{P}_{S}=\bm{P}_{R}(\bm{P}_{R}+\bm{P}_{S})^{+}\bm{P}_{S}
\end{equation}
and it is such that
\begin{equation}\label{eq:130eq}
\bm{P}_{\frown}=2(\bm{P}_{R}:\bm{P}_{S})=2(\bm{P}_{R}(\bm{P}_{R}+\bm{P}_{S})^{+}\bm{P}_{S})
\end{equation}
a projection operator of $\alpha\bm{V}+\beta\bm{W}$ on $\bm{0}$ and of $[\bm{V}, \enspace\bm{W}]$ on $\bm{0}$, as well,  i.e.
\begin{equation}\label{eq:131eq}
\bm{P}_{\frown}(\alpha\bm{V}+\beta\bm{W})=\bm{0}
\end{equation}
\begin{equation}\label{eq:132eq}
\bm{P}_{\frown}[\bm{V}, \enspace\bm{W}]=\bm{0}
\end{equation}
\textbf{Proof}
\\The parallel sum $\bm{P}_{R}:\bm{P}_{S}$ enjoys the property (see e.g.,~\cite{anderson1969series})
\begin{equation}\label{eq:133eq}
\bm{P}_{R}:\bm{P}_{S}=\bm{P}_{R}(\bm{P}_{R}+\bm{P}_{S})^{+}\bm{P}_{S}=\bm{P}_{S}(\bm{P}_{R}+\bm{P}_{S})^{+}\bm{P}_{R}=\bm{P}_{S}:\bm{P}_{R}
\end{equation}
Straightforward computations show that
\[\bm{P}_{\frown}(\alpha\bm{V}+\beta\bm{W})=2(\bm{P}_{R}:\bm{P}_{S})(\alpha\bm{V}+\beta\bm{W})=2\bm{P}_{S}(\bm{P}_{S}+\bm{P}_{R})^{+}\bm{P}_{R}(\alpha\bm{V}+\beta\bm{W})=\]
\begin{equation}\label{eq:134eq}
 =2\alpha\bm{P}_{S}(\bm{P}_{S}+\bm{P}_{R})^{+}\bm{P}_{R}\bm{V}+2\beta\bm{P}_{R}(\bm{P}_{R}+\bm{P}_{S})^{+}\bm{P}_{S}\bm{W}=0\end{equation} 
\[\bm{P}_{\frown}(\bm{V},\,\bm{W})=2(\bm{P}_{R}:\bm{P}_{S})(\bm{V},\,\bm{W})=2\bm{P}_{S}(\bm{P}_{S}+\bm{P}_{R})^{+}\bm{P}_{R}(\bm{V},\,\bm{W})=\]
\begin{equation}\label{eq:135eq}
 =[2\bm{P}_{S}(\bm{P}_{S}+\bm{P}_{R})^{+}\bm{P}_{R}\bm{V}, \enspace 2\bm{P}_{R}(\bm{P}_{R}+\bm{P}_{S})^{+}\bm{P}_{S}\bm{W}]=0\end{equation}
which proves \eqref{eq:131eq} and \eqref{eq:132eq}. 
\begin{flushright}
$\square$
\end{flushright}
The results here below prove useful.
Let $\bm{A}$, $\bm{B}$ and $\bm{C}$ idempotent square matrices of order $n$. Note that for any idempotent matrix $\bm{A}$ a representation of the $\bm{A}=\bm{\Gamma}\bm{\Gamma}^{+}$ holds. As for the idempotent matrix $\bm{A}^{T}$, this representation can be obtain form the rank factorization of $\bm{A}$. Let $\bm{A}=\bm{D}\bm{E}'$, then $\bm{A}^{T}=\bm{I}-\bm{A}\bm{A}^{+}=\bm{I}-\bm{D}\bm{D}^{+}=\bm{\Gamma}\bm{\Gamma}^{+}$, where $\bm{\Gamma}=\bm{D}_{\bot}$.\\
Then, the following statements hold (see ~\cite{berkics2017parallel},~\cite{bernstein2009matrix} p. 201,527,~\cite{piziak1999constructing}, ~\cite{tian2002express} and~\cite{tian2006rank})
\begin{enumerate}
\item \begin{equation}\label{eq:136eq}\alpha(\bm{A}:\bm{B})=(\alpha\bm{A}:\alpha\bm{B}), \enspace \enspace \alpha>0\end{equation}
\item \begin{equation}\label{eq:137eq}(\bm{A}:\bm{B})=-\begin{bmatrix}
\bm{0},&\bm{0},&\bm{I}
\end{bmatrix}\begin{bmatrix}
\bm{A}&\bm{0}&\bm{I}\\
\bm{0}&\bm{B}&\bm{I}\\
\bm{I}&\bm{I}&\bm{0}
\end{bmatrix}
\begin{bmatrix}
\bm{0}\\\bm{0}\\\bm{I}
\end{bmatrix}\end{equation}
\item \textit{The matrix}\begin{equation}\label{eq:138eq}
\bm{P}=2(\bm{A}:\bm{B})    
\end{equation} \textit{is idempotent}
\item \begin{equation}\label{eq:138seq1}
\bm{P}=2\bm{A}(\bm{A}+\bm{B})^{+}\bm{B}=\bm{A}-(\bm{A}-\bm{B}\bm{A})^{+}(\bm{A}-\bm{B}\bm{A})
\end{equation}
\item \begin{equation}\label{eq:138seq}
(\bm{A}:\bm{B}):\bm{C}=\bm{A}:(\bm{B}:\bm{C})
\end{equation}
\item \begin{equation}\label{eq:139eq}
r(\bm{A}:\bm{B})=r(\bm{A})+r(\bm{B})-r(\bm{A}+\bm{B})
\end{equation}
\begin{equation}
=r(\bm{A})+r(\bm{B})-r([\bm{B}\bm{A}^{\top},\enspace\bm{A}]) \enspace \textit{in general}
\end{equation}
\begin{equation}\label{eq:1391eq}
=r(\bm{A})+r(\bm{B})-r([\bm{A}^{\top},\bm{A}])=r(\bm{A})+r(\bm{B})-n, \enspace \textit{if $\bm{B}\bm{A}^{\top}=\bm{A}^{\top}$}
\end{equation}
\textit{as $r([\bm{A}^{\top},\bm{A}])=n$}
\end{enumerate}
Furthermore, thanks to the rank equality (see ~\cite{tian2006rank}) ~\cite{ bernstein2009matrix}),
\begin{equation}\label{eq:140eq}
r(\bm{A}^{\top}+\bm{B}^{\top})=r(\bm{A}+\bm{B})+n-r(\bm{A})-r(\bm{B})\end{equation}
the following holds
\begin{equation}\label{eq:fg}
r(\bm{A}^{\top}:\bm{B}^{\top})=r(\bm{A}^{\top})+r(\bm{B}^{\top})-r(\bm{A}^{\top}+\bm{B}^{\top})=n-r(\bm{A}+\bm{B})=n-r([\bm{\Gamma}, \bm{\Xi}])
\end{equation}
as $r(\bm{A}^{\top})=n-r(\bm{A})$ and $r(\bm{B}^{\top})=n-r(\bm{B})$ and $\bm{A}=\bm{\Gamma}\bm{\Gamma}^{+}$, $\bm{B}=\bm{\Xi}\bm{\Xi}^{+}$.\\
\begin{flushright}
$\square$
\end{flushright}

\section*{Appendix B}\label{sec:appB}
In this Appendix we work out closed-form representations of the principal-part matrices of the Laurent expansion of $\bm{A}^{-1}(z)$ about $z=1$.
To start with, notice that the following holds
\begin{equation}\label{eq:144eq}
\left\{\begin{array}{ll}
\bm{N}_{-m}\bm{A}=\bm{0}\\
\bm{A}\bm{N}_{-m}=\bm{0}\\
\end{array}\right.
\end{equation}
by virtue of \eqref{eq:32eq} and (34) by taking $h=0$.
Solving for $\bm{N}_{-m}$ (see  Lemma 2.3.1 in~\cite{rao1973theory}) yields
\begin{equation}\label{eq:145eq}
\bm{N}_{-m}=\bm{A}^{\bot}\bm{\Phi}_{m}\bm{A}^{\top}=\bm{C}_{0\bot}\bm{C}_{0\bot}^{+}\bm{\Phi}_{m}(\bm{B}_{0\bot}^{'})^{+}\bm{B}_{0\bot}^{'}
\end{equation}
for some $\widetilde{\bm{\Phi}}_{m}$, where $\bm{A}^{\bot}=\bm{I}-\bm{A}^{+}\bm{A}=\bm{C}_{0\bot}\bm{C}_{0\bot}^{+}, \enspace \bm{A}^{\top}=\bm{I}-\bm{A}\bm{A}^{+}=(\bm{B}_{0\bot}^{'})^{+}\bm{B}_{0\bot}^{'}$, (see e.g., ~\cite{faliva2008dynamic}) . This in turn yields \eqref{eq:34eq} by putting
\begin{equation}\label{eq:146eq}
\bm{Z}_{m}=\bm{C}_{0\bot}^{+}\bm{\Phi}_{m}(\bm{B}_{0\bot}^{'})^{+}
\end{equation}
As for $\bm{N}_{-m+1}$, from \eqref{eq:32eq} and (34) we have for $h$=1,
\begin{equation}\label{eq:147eq}
\left\{\begin{array}{ll}
\bm{N}_{-m+1}\bm{A}+\bm{N}_{-m}\bm{A}^{(1)}=\bm{0}\\
\bm{A}\bm{N}_{-m+1}+\bm{A}^{(1)}\bm{N}_{-m}=\bm{0}\\
\end{array}\right.
\end{equation}
Solving for $\bm{N}_{-m+1}$ (see Theorem 2.3.3, formula (2.3.7) in~\cite{rao1973theory}) yields 
\begin{equation}\label{eq:148eq}
\bm{N}_{-m+1}=-\bm{A}^{+}\bm{A}^{(1)}\bm{N}_{-m}-\bm{N}_{-m}\bm{A}^{(1)}\bm{A}^{+}+\bm{C}_{0\bot}\bm{S}_{1,1}\bm{B}_{0\bot}^{'}
\end{equation}
for some $\bm{S}_{1,1}$, as $\bm{N}_{-m}\bm{A}^{\top}=\bm{N}_{-m}$ in light of \eqref{eq:144eq}. \\
Now, let $m$>2. Then from  \eqref{eq:32eq} and (34) we have for $h$=2
\begin{equation}\label{eq:149eq}
\left\{\begin{array}{ll}
\bm{N}_{-m+2}\bm{A}+\bm{N}_{-m+1}\bm{A}^{(1)}+\frac{1}{2}\bm{N}_{-m}\bm{A}^{(2)}=\bm{0}\\
\bm{A}\bm{N}_{-m+2}+\bm{A}^{(1)}\bm{N}_{-m+1}+\frac{1}{2}\bm{A}^{(2)}\bm{N}_{-m}=\bm{0}\\
\end{array}\right.
\end{equation}
Pre and post-multiply the first equation by $\bm{C}_{0\bot}^{+}$ and $\bm{C}_{0\bot}$ and the second one by $\bm{B}_{0\bot}^{'}$ and $(\bm{B}_{0\bot}^{'})^{+}$, respectively. Then, by making use of \eqref{eq:148eq}, a simple computation yields 
\begin{equation}\label{eq:150eq}
\left\{\begin{array}{ll}
\bm{S}_{1,1}\bm{K}_{1}=-(\bm{C}_{0\bot})^{+}\bm{N}_{-m}\bm{A}^{[2]}\bm{C}_{0\bot}\\
\bm{K}_{1}\bm{S}_{1,1}=-\bm{B}^{'}_{0\bot}\bm{A}^{[2]}\bm{N}_{-m}(\bm{B}^{'}_{0\bot})^{+}\\
\end{array}\right.
\end{equation}
where $\bm{K}_{1}$ and $\bm{A}^{[2]}$ are defined in \eqref{eq:38eq} and \eqref{eq:8eq}.\\ Solving for $\bm{S}_{1,1}$ and pre and post multiplying the result by $\bm{C}_{0\bot}$ and $\bm{B}_{0\bot}^{'}$ gives
\begin{equation}\label{eq:151eq}
\bm{C}_{0\bot}\bm{S}_{1,1}\bm{B}_{0\bot}^{'}=-\bm{\Theta}_{1}\bm{A}^{[2]}\bm{N}_{-m}-\bm{N}_{-m}\bm{A}^{[2]}\bm{\Theta}_{1}+\bm{C}_{0\bot}\bm{C}_{1\bot}\bm{S}_{1,2}\bm{B}_{1\bot}^{'}\bm{B}_{0\bot}^{'}
\end{equation}
for some $\bm{S}_{1,2}$ and $\bm{\Theta}_{1}$ given by \eqref{eq:12eq}, as $\bm{N}_{-m}(\bm{B}_{0\bot}^{'})^{+}\bm{K}_{1}^{\top}\bm{B}_{0\bot}^{'}=\bm{K}_{-m}$ and $(\bm{C}_{0\bot})^{+}\bm{C}_{0\bot}\bm{N}_{-m}=\bm{N}_{-m}$. \\
Accordingly,the representation of $\bm{N}_{-m+1}$ for $m>2$ becomes
\begin{equation}\label{eq:148eqzl}
\bm{N}_{-m+1}=-\bm{A}^{+}\bm{A}^{(1)}\bm{N}_{-m}-\bm{N}_{-m}\bm{A}^{(1)}\bm{A}^{+}-\bm{\Theta}_{1}\bm{A}^{[2]}\bm{N}_{-m}-\bm{N}_{-m}\bm{A}^{[2]}\bm{\Theta}_{1}+\bm{C}_{0\bot}\bm{C}_{1\bot}\bm{S}_{1,2}\bm{B}_{1\bot}^{'}\bm{B}_{0\bot}^{'}
\end{equation}
At this point, let us move to $\bm{N}_{-m+2}$. Solving \eqref{eq:149eq} for $\bm{N}_{-m+2}$ yields
\begin{equation*}
\bm{N}_{-m+2}=-\bm{A}^{+}\bm{A}^{(1)}\bm{N}_{-m+1}-\frac{1}{2}\bm{A}^{+}\bm{A}^{(2)}\bm{N}_{-m}-\bm{N}_{-m+1}\bm{A}^{(1)}\bm{A}^{+}-\frac{1}{2}\bm{N}_{-m}\bm{A}^{(2)}\bm{A}^{+}+
\end{equation*}
\begin{equation}\label{eq:152eq}
+\bm{A}^{+}\bm{A}\bm{N}_{-m+1}\bm{A}^{(1)}\bm{A}^{+}+
\bm{C}_{0\bot}\bm{S}_{2,1}\bm{B}_{0\bot}^{'}
\end{equation}
which can be also written as
\begin{equation*}
\bm{N}_{-m+2}=-\bm{A}^{+}\bm{A}^{(1)}\bm{N}_{-m+1}\bm{A}^{\top}-\bm{A}^{+}\bm{A}^{[2]}\bm{N}_{-m}-\bm{A}^{+}\bm{A}^{(1)}\bm{A}^{+}\bm{A}^{(1)}\bm{N}_{-m}-\bm{N}_{-m+1}\bm{A}^{(1)}\bm{A}^{+}+
\end{equation*}
\begin{equation}\label{eq:tt}
 -\bm{N}_{-m}\bm{A}^{[2]}\bm{A}^{+}-\bm{N}_{-m}\bm{A}^{(1)}\bm{A}^{+}\bm{A}^{(1)}\bm{A}^{+}+\bm{C}_{0\bot}\bm{S}_{2,1}\bm{B}^{'}_{0\bot}   
\end{equation}
for some $\bm{S}_{2,1}$. Formula \eqref{eq:tt} can be expressed in term of the leading principal-part matrix $\bm{N}_{-m}$ as follows
\[\bm{N}_{-m+2}=\bm{A}^{+}\bm{A}^{(1)}\bm{N}_{-m}\bm{A}^{(1)}\bm{A}^{+} -\bm{A}^{+}\bm{A}^{[2]}\bm{N}_{-m}  -\bm{N}_{-m}\bm{A}^{[2]}\bm{A}^{+}+\]\begin{equation}\label{eq:41eq}
-\bm{C}_{0\bot}\bm{S}_{1,1}\bm{B}_{0\bot}^{'}\bm{A}^{+}\bm{A}^{(1)}-\bm{A}^{+}\bm{A}^{(1)}
\bm{C}_{0\bot}\bm{S}_{1,1}\bm{B}_{0\bot}^{'}+
\bm{C}_{0\bot}\bm{S}_{2,1}\bm{B}_{0\bot}^{'}
\end{equation}
Coming back to $\bm{N}_{-m+1}$, let $m$>3. Then, from \eqref{eq:32eq} and (34) we have for $h$=3 
\begin{equation}\label{eq:153eq}
\begin{split}
&\bm{N}_{-m+3}\bm{A}+\bm{N}_{-m+2}\bm{A}^{(1)}+\frac{1}{2}\bm{N}_{-m+1}\bm{A}^{(2)}+\frac{1}{6}\bm{N}_{-m}\bm{A}^{(3)}=\bm{0}\\
&\bm{A}\bm{N}_{-m+3}+\bm{A}^{(1)}\bm{N}_{-m+2}\frac{1}{2}\bm{A}^{(2)}\bm{N}_{-m+1}\frac{1}{6}\bm{A}^{(3)}\bm{N}_{-m}=\bm{0}
\end{split}\end{equation}
Pre and post-multiply the first equation by $\bm{C}_{1\bot}^{+}\bm{C}_{0\bot}^{+}$ and $\bm{C}_{0\bot}\bm{C}_{1\bot}$ respectively, and the second one by $\bm{B}_{1\bot}^{'}\bm{B}_{0\bot}^{'}$ and $(\bm{B}_{0\bot}^{'})^{+}(\bm{B}_{1\bot}^{'})^{+}$, respectively. Then, by making use of \eqref{eq:152eq}, \eqref{eq:148eq} and \eqref{eq:151eq}, a simple computation yields the  system
\begin{equation}\label{eq:153seq}
\left\{\begin{array}{ll}
\bm{S}_{1,2}\bm{K}_{2}=-\bm{C}_{1\bot}^{+}\bm{C}_{0\bot}^{+}\bm{N}_{-m}\bm{A}^{[3]}\bm{C}_{0\bot}\bm{C}_{1\bot}\\
\bm{K}_{2}\bm{S}_{1,2}=-\bm{B}^{'}_{1\bot}\bm{B}^{'}_{0\bot}\bm{A}^{[3]}\bm{N}_{-m}(\bm{B}^{'}_{0\bot})^{+}(\bm{B}^{'}_{1\bot})^{+}\\
\end{array}\right.
\end{equation}
as $\bm{C}_{1\bot}^{+}\bm{C}_{0\bot}^{+}\bm{\Theta}_{1}=\bm{0}$ and $\bm{\Theta}_{1}(\bm{B}_{0\bot}^{'})^{+}(\bm{B}_{1\bot}^{'})^{+}=\bm{0}$ with $\bm{K}_{2}$ and $\bm{A}^{[3]}$ as defined in \eqref{eq:39eq} and \eqref{eq:9eq}.
Solving for $\bm{S}_{1,2}$ and pre and post-multiplying by $\bm{C}_{0\bot}\bm{C}_{1\bot}
$ and $\bm{B}_{1\bot}^{'}\bm{B}_{0\bot}^{'}$ yields
\begin{equation}\label{eq:154eq}
\bm{C}_{0\bot}\bm{C}_{1\bot}\bm{S}_{1,2}\bm{B}_{1\bot}^{'}\bm{B}_{0\bot}^{'}=-\bm{\Theta}_{2}\bm{A}^{[3]}\bm{N}_{-m}-\bm{N}_{-m}\bm{A}^{[3]}\bm{\Theta}_{2}+\bm{C}_{0\bot}\bm{C}_{1\bot}\bm{C}_{2\bot}\bm{S}_{1,3}\bm{B}_{2\bot}^{'}\bm{B}_{1\bot}^{'}\bm{B}_{0\bot}^{'}
\end{equation}
for some $\bm{S}_{1,2}$ and $\bm{\Theta}_{2}$ given by \eqref{eq:13eq}, as $\bm{N}_{-m}(\bm{B}_{0\bot}^{'})^{+}(\bm{B}_{1\bot}^{'})^{+}(\bm{I}-\bm{K}_{2}\bm{K}_{2}^{+})\bm{B}_{1\bot}^{'}\bm{B}_{0\bot}^{'}=\bm{N}_{-m}$.\\
Accordingly, $\bm{N}_{-m+1}$ for $m>3$ can be also expressed as
\begin{equation*}
\bm{N}_{-m+1}=-\bm{A}^{+}\bm{A}^{(1)}\bm{N}_{-m}-\bm{N}_{-m}\bm{A}^{(1)}\bm{A}^{+}-\bm{\Theta}_{1}\bm{A}^{[2]}\bm{N}_{-m}-\bm{N}_{-m}\bm{A}^{[2]}\bm{\Theta}_{1}-\bm{\Theta}_{2}\bm{A}^{[3]}\bm{N}_{-m}+
\end{equation*}
\begin{equation}\label{eq:148eqzt}
-\bm{N}_{-m}\bm{A}^{[3]}\bm{\Theta}_{2}+\bm{C}_{0\bot}\bm{C}_{1\bot}\bm{C}_{2\bot}\bm{S}_{1,3}\bm{B}_{2\bot}^{'}\bm{B}_{1\bot}^{'}\bm{B}_{0\bot}^{'}
\end{equation}
Coming back to $\bm{N}_{-m+2}$, let $m$> 3 and refer to the system  \eqref{eq:153eq}. Pre and post-multiplying the first equation by $\bm{C}_{0\bot}^{+}$ and $\bm{C}_{0\bot}$, and the second equation by $\bm{B}_{0\bot}^{'}$ and $(\bm{B}_{0\bot}^{'})^{+}$, gives 
\begin{equation}\label{eq:155eq}
\left\{\begin{array}{ll}
\bm{S}_{2,1}\bm{K}_{1}=-(\bm{C}_{0\bot})^{+}\bm{N}_{-m+1}\bm{A}^{[2]}\bm{C}_{0\bot}-(\bm{C}_{0\bot})^{+}\bm{N}_{-m}\breve{\bm{A}}^{[3]}\bm{C}_{0\bot}\\
\bm{K}_{1}\bm{S}_{2,1}=-\bm{B}^{'}_{0\bot}\bm{A}^{[2]}\bm{N}_{-m+1}(\bm{B}^{'}_{0\bot})^{+}-\bm{B}^{'}_{0\bot}\dot{\bm{A}}^{[3]}\bm{N}_{-m}(\bm{B}^{'}_{0\bot})^{+}\\
\end{array}\right.
\end{equation}
where 
\begin{equation}\label{eq:156eq}
\breve{\bm{A}}^{[3]}=(\frac{1}{6}\bm{A}^{(3)}-\frac{1}{2}\bm{A}^{(2)}\bm{A}^{+}\bm{A}^{(1)})=\bm{A}^{[3]}+\bm{A}^{(1)}\bm{A}^{+}\bm{A}^{[2]}+\bm{A}^{[2]}\bm{\Theta}_{1}\bm{A}^{[2]}
\end{equation}
\begin{equation}\label{eq:157eq}
\dot{\bm{A}}^{[3]}=(\frac{1}{6}\bm{A}^{(3)}-\frac{1}{2}\bm{A}^{(1)}\bm{A}^{+}\bm{A}^{(2)})=\bm{A}^{[3]}+\bm{A}^{[2]}\bm{A}^{+}\bm{A}^{(1)}+\bm{A}^{[2]}\bm{\Theta}_{1}\bm{A}^{[2]}
\end{equation}
Solving for $\bm{S}_{2,1}$ and pre and post-multiplying by $\bm{C}_{0\bot}$ and $\bm{B}_{0\bot}^{'}$, gives 
\begin{equation*}
\bm{C}_{0\bot}\bm{S}_{2,1}\bm{B}^{'}_{0\bot}=-\bm{\Theta}_{1}\bm{A}^{[2]}\bm{N}_{-m+1}(\bm{B}^{'}_{0\bot})^{+}\bm{K}^{\top}_{1}\bm{B}^{'}_{0\bot}-\bm{\Theta}_{1}\dot{\bm{A}}^{[3]}\bm{N}_{-m}+
\end{equation*}
\begin{equation}\label{eq:158eq}
-\bm{C}_{0\bot}\bm{C}^{+}_{0\bot}\bm{N}_{-m+1}\bm{A}^{[2]}\bm{\Theta}_{1}-\bm{N}_{-m}\breve{\bm{A}}^{[3]}\bm{\Theta}_{1}+\bm{C}_{0\bot}\bm{C}_{1\bot}\bm{S}_{2,2}\bm{B}^{'}_{1\bot}\bm{B}^{'}_{0\bot}\end{equation}
\\for some $\bm{S}_{2,2}$, as
$\bm{N}_{-m}(\bm{B}_{0\bot}^{'})^{+}\bm{K}_{1}^{\top}\bm{B}_{0\bot}^{'}=\bm{N}_{-m}$ and $\bm{C}_{0\bot}\bm{C}_{0\bot}^{+}\bm{N}_{-m}=\bm{N}_{-m}.$ \\
Accordingly, the representation of $\bm{N}_{-m+2}$ when $m>3$ is
\[
\bm{N}_{-m+2}=-\bm{A}^{+}\bm{A}^{(1)}\bm{N}_{-m+1}\bm{A}^{\top}-\bm{A}^{+}\bm{A}^{[2]}\bm{N}_{-m}-\bm{A}^{+}\bm{A}^{(1)}\bm{A}^{+}\bm{A}^{(1)}\bm{N}_{-m}-\bm{N}_{-m+1}\bm{A}^{(1)}\bm{A}^{+}+
\]
\[-\bm{N}_{-m}\bm{A}^{[2]}\bm{A}^{+}-\bm{N}_{-m}\bm{A}^{(1)}\bm{A}^{+}\bm{A}^{(1)}\bm{A}^{+}-\bm{\Theta}_{1}\bm{A}^{[2]}\bm{N}_{-m+1}(\bm{B}^{'}_{0\bot})^{+}\bm{K}^{\top}_{1}\bm{B}^{'}_{0\bot}-\bm{\Theta}_{1}\dot{\bm{A}}^{[3]}\bm{N}_{-m}+ \]
\begin{equation}\label{eq:ttz}
-\bm{C}_{0\bot}\bm{C}^{+}_{0\bot}\bm{N}_{-m+1}\bm{A}^{[2]}\bm{\Theta}_{1}-\bm{N}_{-m}\breve{\bm{A}}^{[3]}\bm{\Theta}_{1}+\bm{C}_{0\bot}\bm{C}_{1\bot}\bm{S}_{2,2}\bm{B}^{'}_{1\bot}\bm{B}^{'}_{0\bot}
\end{equation}
Finally, as for $\bm{N}_{-m+3}$, let $m$>3 and refer to system  \eqref{eq:153eq}. Solving for $\bm{N}_{-m+3}$ yields 
\begin{equation*}
\bm{N}_{-m+3}=-\bm{A}^{+}\bm{A}^{(1)}\bm{N}_{-m+2}-\frac{1}{2}\bm{A}^{+}\bm{A}^{(2)}\bm{N}_{-m+1}-\frac{1}{6}\bm{A}^{+}\bm{A}^{(3)}\bm{N}_{-m}-\bm{N}_{-m+2}\bm{A}^{(1)}\bm{A}^{+}+
\end{equation*}
\begin{equation*}
\frac{1}{2}
\bm{N}_{-m+1}\bm{A}^{(2)}\bm{A}^{+} -\frac{1}{6}\bm{N}_{-m}\bm{A}^{(3)}\bm{A}^{+}+\bm{A}^{+}\bm{A}\bm{N}_{-m+2}\bm{A}^{(1)}\bm{A}^{+}+\frac{1}{2}\bm{A}^{+}\bm{A}\bm{N}_{-m+1}\bm{A}^{(2)}\bm{A}^{+}+
\end{equation*}
\begin{equation}\label{eq:159eq}
+\bm{C}_{0 \bot}\bm{S}_{3,1}\bm{B}_{0 \bot}^{'}
\end{equation}
which can be alternatively expressed as follows
\begin{center}
$\bm{N}_{-m+3}=-\bm{A}^{+}\bm{A}^{(1)}\bm{N}_{-m+2}\bm{A}^{\top}-\bm{A}^{+}\bm{A}^{[2]}\bm{N}_{-m+1}\bm{A}^{\top}-\bm{A}^{+}\widetilde{\bm{A}}^{(2)}\bm{N}_{-m+1}\bm{A}^{\top}-\bm{A}^{+}\bm{A}^{[3]}\bm{N}_{-m}+$\end{center} 
\begin{center} $-\bm{A}^{+}\widetilde{\bm{A}}^{(3)}\bm{N}_{-m}-\bm{N}_{-m+2}\bm{A}^{(1)}\bm{A}^{+}-\bm{N}_{-m+1}\bm{A}^{[2]}\bm{A}^{+}-\bm{N}_{-m+1}\widetilde{\bm{A}}^{(2)}\bm{A}^{+}+$\end{center}
\begin{equation}\label{eq:43eq}
-\bm{N}_{-m}\bm{A}^{[3]}\bm{A}^{+}-\bm{N}_{-m}\widetilde{\bm{A}}^{(3)}\bm{A}^{+}+\bm{C}_{0\bot}\bm{S}_{3,1}\bm{B}_{0\bot}^{'},
\end{equation}\vspace{0.01mm}
or in term of the leading principal-part matrix $\bm{N}_{m}$ as 
\begin{center}
$\bm{N}_{-m+3}=-\bm{A}^{+}\bm{A}^{[3]}\bm{N}_{-m}-\bm{N}_{-m}\bm{A}^{[3]}\bm{A}^{+}+\bm{A}^{+}\bm{A}^{(1)}\bm{N}_{-m}\bm{A}^{[2]}\bm{A}^{+}+\bm{A}^{+}\bm{A}^{[2]}\bm{N}_{-m}\bm{A}^{(1)}\bm{A}^{+}+ $\end{center} 
\begin{center} 
$
-\bm{A}^{+}\bm{A}^{[2]}\bm{\Theta}_{1}\bm{A}^{[2]}\bm{N}_{-m}-\bm{N}_{-m}\bm{A}^{[2]}\bm{\Theta}_{1}\bm{A}^{[2]}\bm{A}^{+}-\bm{A}^{+}\bm{A}^{(1)}\bm{C}_{0\bot}\bm{S}_{2,1}\bm{B}_{0\bot}^{'} +
$\end{center}
\begin{center} $
+\bm{A}^{+}\bm{A}^{(1)}\bm{C}_{0\bot}\bm{S}_{1,1}\bm{B}_{0\bot}^{'}\bm{A}^{(1)}\bm{A}^{+}-\bm{C}_{0\bot}\bm{S}_{1,1}\bm{B}_{0\bot}^{'}\bm{A}^{[2]}\bm{A}^{+}-\bm{A}^{+}\bm{A}^{[2]}\bm{C}_{0\bot}\bm{S}_{1,1}\bm{B}_{0\bot}^{'}+ $\end{center}
\begin{equation}\label{eq:43seq}
-\bm{C}_{0\bot}\bm{S}_{2,1}\bm{B}_{0\bot}^{'}\bm{A}^{(1)}\bm{A}^{+} +\bm{C}_{0\bot}\bm{S}_{3,1}\bm{B}_{0\bot}^{'}
\end{equation}
\begin{flushright}
$\square$
\end{flushright}

\bibliographystyle{apalike}
\bibliography{cointegration}

\end{document}